\documentclass{article}

     \usepackage[preprint]{neurips_2019}

\usepackage[utf8]{inputenc} % allow utf-8 input
\usepackage[T1]{fontenc}    % use 8-bit T1 fonts
\usepackage{hyperref}       % hyperlinks
\usepackage{url}            % simple URL typesetting
\usepackage{booktabs}       % professional-quality tables
\usepackage{amsfonts}       % blackboard math symbols
\usepackage{nicefrac}       % compact symbols for 1/2, etc.
\usepackage{microtype}      % microtypography

\usepackage{amssymb}
\usepackage{amsmath}
\usepackage{bm}
\usepackage{algorithm}
\usepackage{algpseudocode}
\usepackage{comment}
\usepackage{graphicx}

\title{Selective Particle Attention: Visual Feature-Based Attention in Deep Reinforcement Learning}

\author{%
  Sam Blakeman\\
  Centre for Brain and Cognitive Development \\
  Department of Psychological Sciences\\
  Birkbeck, University of London \\
  Malet Street, WC1E 7HX UK \\
  \texttt{sblake03@mail.bbk.ac.uk} \\
  \And
  Denis Mareschal \\
  Centre for Brain and Cognitive Development \\
  Department of Psychological Sciences\\
  Birkbeck, University of London \\
  Malet Street, WC1E 7HX UK \\
  \texttt{d.mareschal@bbk.ac.uk} \\
}

\begin{document}

\maketitle

\begin{abstract}
The human brain uses selective attention to filter perceptual input so that only the components that are useful for behaviour are processed using its limited computational resources. This process is thought to originate in the Pre-Frontal Cortex (PFC) and can be found across several perceptual modalities including vision. We focus on one particular form of visual attention known as feature-based attention, which is concerned with identifying features of the visual input that are important for the current task regardless of their spatial location. Visual feature-based attention has been proposed to improve the efficiency of Reinforcement Learning (RL) by reducing the dimensionality of state representations and guiding learning towards relevant features. Despite achieving human level performance in complex perceptual-motor tasks, Deep RL algorithms have been consistently criticised for their poor efficiency and lack of flexibility. Visual feature-based attention therefore represents one option for addressing these criticisms. Nevertheless, it is still an open question how the brain is able to learn which features to attend to during RL. To help answer this question we propose a novel algorithm, termed \textit{Selective Particle Attention (SPA)}, which imbues a Deep RL agent with the ability to perform selective feature-based attention. SPA uses a particle filter to select combinations of features produced by a pre-trained deep convolutional neural network. The selected features are then passed on to a Deep RL algorithm for action selection. This architecture mimics the problem faced by the brain; given existing perceptual features, which ones are useful for the current task? SPA learns which combinations of features to attend to based on their bottom-up saliency and how accurately they predict future reward. We evaluate SPA on a multiple choice task and a 2D video game that both involve raw pixel input and dynamic changes to the task structure. We show various benefits of SPA over approaches that naively attend to either all or random subsets of features. Our results demonstrate (1) how visual feature-based attention in Deep RL models can improve their learning efficiency and ability to deal with sudden changes in task structure and (2) that particle filters may represent a viable computational account of how visual feature-based attention occurs in the brain.
\end{abstract}

\section{Introduction}

The human brain is constantly faced with a multi-sensory, high-dimensional stream of perceptual input that has to be processed in order to select actions. Critical to this process is the identification of features of the environment that are relevant for the task at hand. Much of the perceptual input is uninformative given the current goal and so its removal can greatly simplify the learning process. The process of selecting the relevant sensory features for the current task is often referred to as selective attention and is it thought to be the responsibility of the Pre-Frontal Cortex (PFC) \citep{miller2001integrative}. 

Selective attention can occur across several modalities including vision and audition. In vision, selective attention is often broadly split into two types; spatial and feature-based attention \citep{lindsay2020attention}. Spatial attention refers to the selective processing of specific areas of the visual field. In comparison, feature-based attention is used to selectively process specific features of the visual input regardless of their location in the visual field and is the focus of this paper. Feature-based attention is typically studied by priming individuals to attend to a certain feature and then measuring their ability to detect the primed feature. For example, participants who are primed to attend to a specific orientation of visual grating are subsequently better at detecting that grating \citep{rossi1995feature}. From a neural point of view, the firing of neurons that encode the attended feature appear to increase, while the firing rates of neurons encoding the non-attended feature appear to decrease \citep{treue1999feature, saenz2002global}. This modulation of firing rates in the visual stream is thought to originate from the PFC \citep{bichot2015source,paneri2017top} and be most effective in higher order visual areas \citep{lindsay2018biological}. Importantly, these top-down goal-driven influences from the PFC are thought to work in combination with bottom-up influences, which are driven by intrinsic properties of the stimuli that are task-agnostic \citep{wolfe2004attributes}. While these priming studies demonstrate the effect of selective attention they do not tackle the problem of how to learn which features to attend to. 

The work in this paper focuses on a computational account of how visual feature-based attention can interface with Reinforcement Learning (RL) to help us learn which features to attend to in a task with a specified reward structure. Selective feature-based attention can greatly reduce the complexity of the RL problem by reducing the dimensionality of the state representation to only the features that are important for the current task \citep{jones2010integrating, niv2015reinforcement, wilson2012inferring}. It is typically thought to be much faster acting than the incremental learning of new representations as it allows for the quick and flexible re-purposing of existing representations. Selective attention in RL can affect both learning and choice. During learning, selective attention can modify the magnitude of weight updates for each feature and during choice it can alter the magnitude of each features contribution to a decision \citep{leong2017dynamic}. There is therefore a reciprocal relationship between attention and learning in RL; attention biases learning but learning guides which features are attended to. It remains an open question how we learn which features to attend to based on the reward structure of the current task. One proposal is that we learn to attend to the features that are most predictive of reward \citep{mackintosh1975theory}. For example, people are able to switch between an object-based or a feature-based state representation based on which one is the best predictor of reward \citep{farashahi2017feature}.

Advances in Deep RL have provided the opportunity to begin examining how the brain may go from naturalistic high-dimensional input to action based on reward signals. Deep RL has been particularly powerful in the visual domain where Deep RL agents have learnt to perform complex tasks from raw pixel inputs, such as playing video games at human level performance \citep{mnih2015human, mnih2016asynchronous}. These models typically rely on the use of Deep Convolutional Neural Networks (DCNNs) to approximate a value function and/or a policy. This is of interest to cognitive scientists and neuroscientists because a growing body of evidence is suggesting that the hierarchical representations learnt by DCNNs are similar to those found in the ventral stream of the human brain \citep{gucclu2015deep, yamins2016using, schrimpf2018brain}. Despite these successes, one consistent criticism of Deep RL models is that they lack the efficiency and flexibility demonstrated by human learning. Selective attention represents one potential mechanism for helping to address these issues. Selective attention mechanisms have been applied to DNNs with great success, helping to focus in on both spatial and feature-based components of visual input \citep{mnih2014recurrent, fu2017look, xiao2015application, ba2014multiple, xu2015show}. Within the domain of RL, selective attention to visual input has been applied to DNNs in the form of self-attention, which maps query and key-pair vectors to an output vector \citep{manchin2019reinforcement, mott2019towards, zambaldi2018relational, bramlage2020attention}. This computation is fully differentiable and so can be trained end-to-end using backpropagation.

These forms of selective attention in Deep RL, and indeed other forms \citep{sorokin2015deep}, have several striking differences to feature-based attention in the human visual cortex. Firstly, when presented with a novel situation humans typically select from pre-existing features and use a function of these features to guide decision-making, rather than learning a completely new set of features. This is important because it eludes to one of the most powerful properties of selective feature-based attention; it can quickly and dynamically re-purpose representations without over-writing them. In comparison, learning new features in the visual stream is thought to occur slowly over many experiences and is often referred to as perceptual learning \citep{schyns1998development}. Being fully differentiable, the aforementioned approaches rely on a selective attention mechanism that ultimately learns the underlying representations that are to be attended to, which is both slow and inflexible. A second noticeable difference between current visual selective attention mechanisms in Deep RL and those in the brain is the lack of bottom-up influences. Bottom-up influences are a critical component of human visual attention and help to guide us to salient pieces of information in the environment \citep{treue2003visual}. This suggests that their inclusion in Deep RL models may help to improve their performance and bring them closer to models of human visual selective attention.

With these criticisms in mind, we propose a novel algorithm that we term \textit{Selective Particle Attention ($SPA$)}, which implements visual feature-based attention in a Deep RL agent. At its core $SPA$ uses Bayesian principles to quickly and flexibly re-purposes features that are learnt slowly over many examples. This speeds up learning by reducing the dimensionality of the problem and reduces interference by preserving the underlying features. The approach resembles the problem faced by the human brain in several important ways. Firstly, we use a pre-trained DCNN to extract features of the visual input. In particular we use VGG-16 due to its relatively low computational costs and good correspondence with representations found in the visual stream \citep{schrimpf2018brain}. Importantly, no further training of VGG-16 is performed in order to mimic how the human brain has to pick from existing representations that change very slowly in sensory cortices. Once the features have been extracted from VGG-16, we use a particle filter to implement selective feature-based attention. The goal of the particle filter is to select combinations of features that are useful for predicting reward given the current task, thereby reducing the dimensionality of the problem. This is analogous to the PFC, which implements a mechanism that selectively attends to specific features of the visual input given the current goal \citep{miller2001integrative}. In addition, particle filters have been proposed to represent a computationally plausible model of selective attention in the brain \citep{radulescu2019holistic} and have been shown to better model shifts in attention than gradual error-driven learning \citep{radulescuparticle}. After the particle filter has been used to selectively attend to specific visual features, we apply a Deep RL algorithm to approximate the value function and/or policy. This can be tentatively compared to the role of the striatum, which is thought to evaluate states and/or actions based on inputs from cortical regions \citep{schultz1998predictive, houk199513, joel2002actor, maia2009reinforcement, setlow2003neural} and temporal difference errors \citep{schultz1997neural}. 

We assess our approach, on two key tasks; a multiple choice task and a 2D video game based on collecting objects. Both tasks involve processing observations from raw pixel input and dealing with unannounced changes in task structure. In both cases the selective attention mechanism of $SPA$ led to improved performance in terms of the onset, speed, robustness and flexibility of learning compared to naive approaches that either attended to all or a random subset of features. We also show that these findings occur independently of the RL algorithm used, making it applicable to a variety of problems. Overall our results demonstrate that $SPA$ is a viable method for performing visual feature-based attention in a Deep RL agent and that it may capture some of the key computational properties of selective attention in the brain. In particular, $SPA$ provides a mechanistic explanation of how bottom-up and top-down attention may interact in the brain in order to guide feature selection based on the task at hand.

\section{Methods} \label{SA_methods}

\begin{figure}
	\centering
	\includegraphics[height=9cm, width=14cm]{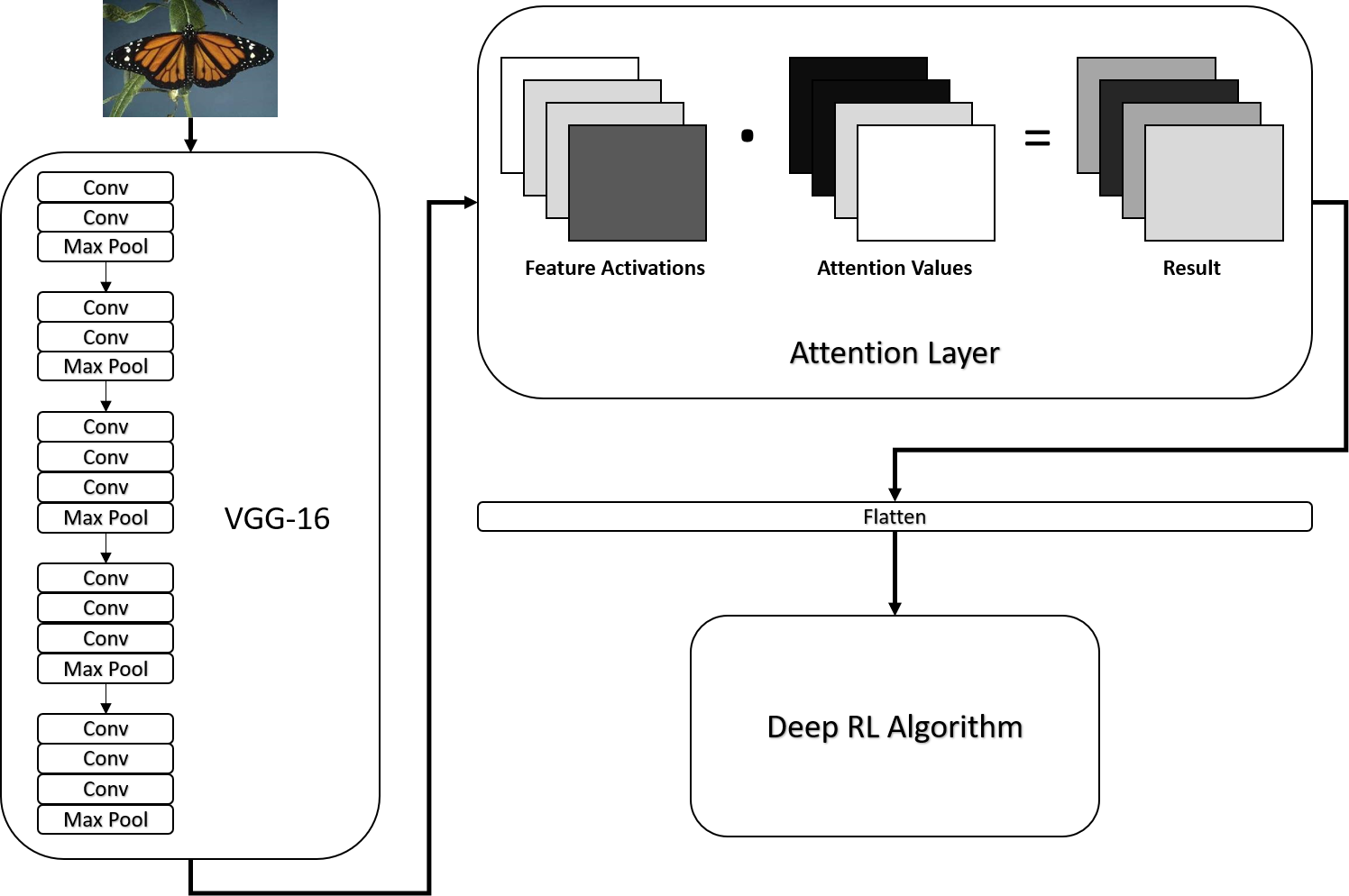}
	\caption{\textit{Architecture of Selective Particle Attention (SPA). The pre-trained deep convolutional neural network VGG-16 is used to extract 2D feature maps from an image. The feature maps are multiplied by their corresponding attention values in an element-wise fashion. The result is then reshaped into a single vector and passed to a deep reinforcement learning algorithm for action selection.}}
	\label{fig:SA_architecture}
\end{figure}

Our model consists of three main components; a pre-trained Deep Convolutional Neural Network (DCNN) for feature extraction, a particle filter for selective attention and a Deep RL algorithm for action selection. Figure \ref{fig:SA_architecture} shows a diagram of the overall agent architecture. The subsequent sections break-down each component and their underlying mechanisms in more detail.

\subsection{VGG-16}

To perform feature extraction on raw pixel values we use a pre-trained Deep Convolutional Neural Network (DCNN). We chose VGG-16 \citep{simonyan2014very} as the DCNN for all simulations because of its relative simplicity and good correspondence with representations found in the visual stream \citep{schrimpf2018brain}. Figure \ref{fig:SA_architecture} shows the architecture of VGG-16, which consists of 5 blocks of convolutional layers each with a max pooling layer at the end. VGG-16 was trained on a classification task using the Imagenet dataset \citep{deng2009imagenet}, which consists of naturalistic images. For our simulations we performed no further training on VGG-16 and removed the fully connected layers after the final convolutional block. This left us with 512 feature maps as output, which served as the basis for selective attention. For all experiments the input images were resized to be 224 X 224 pixels in order to match the required inputs of VGG-16. Pre-processing was performed in the standard manner for VGG-16 \citep{simonyan2014very}.

\subsection{Attention Layer}

The attention layer is applied to the final feature maps provided by VGG-16 using a process similar to the one described by \citet{luo2020costs}. Attention is represented as a $K$ dimensional vector that can take on any real valued number between 0 and 1 inclusive:

\begin{equation}
    \mathbf{A} \in [0, 1]^K
\end{equation}

Where $K$ is the number of feature maps, which in our case was 512. This attention vector is applied to the final feature maps of VGG-16 using the hadamard product between $\mathbf{A}$ and the values of each feature map. This requires that each entry in $\mathbf{A}$ is replicated to match the dimensions of a single feature map, with the same attentional value applied across all spatial locations. This process re-weights the feature map activations, amplifying feature maps with a large attentional weight. The output of the attention layer is then reshaped into a single vector and passed on to a Deep Reinforcement Learning (RL) algorithm for action selection. Our approach is independent of the Deep RL algorithm used, so long as it involves the approximation of a state value function (See Section \ref{particle_filter}).

\subsection{Particle Filter} \label{particle_filter}

The particle filter is responsible for learning the values of the attention vector given the current task. Particle filters are typically used to estimate the value of a latent variable ($X$) given noisy samples of an observed variable ($O$) when the number of potential values is large. The overall goal of a particle filter is to use a set of 'particles' to represent a posterior distribution over the latent variable. Each particle represents a belief or hypothesis about the value of the latent variable and the density of the particles can be used to approximate the posterior distribution.

In our case the latent variable $X$ represents the configuration of features that are useful for the current task. We use a value of 1 to denote a feature as useful and a value of 0 to denote a feature as not useful. This means that $x$ can be any binary vector of size $K$: 

\begin{equation}
    x \in \{0, 1\}^K
\end{equation}

Where $K$ is the number of features, which in our case was 512 corresponding to the number of feature maps of VGG-16. As the number of possible values that $X$ can take is $2^K$, we use $N$ particles to approximate the posterior distribution over $X$ where $N << 2^K$. The state $x^i$ of the $i^{th}$ particle therefore corresponds to a binary vector of length $K$ and provides a hypothesis about which features it deems relevant for the current task.

A particle filter consists of two main steps; a movement step and an observation step. In the movement step, the particles are updated based on some known transition probability for the latent variable:

\begin{equation}
    x' \sim P(X'|x) \label{eq:movement}
\end{equation}

This process is often used to represent the passing of time. In our approach we introduce the notion of bottom-up attention during the movement step. Let $\overline{f}^k_t$ denote the average value over all the units in feature map $k$ from VGG-16 at time $t$. At each time-step $t$ a particle is updated as follows with some probability $\phi$:

\begin{align}
    v^k_t &= \frac{\overline{f}^k_t}{\sum^K_{j=1}\overline{f}^j_t} \\
    p^k &= \frac{\exp({v^k_t * \tau_{\text{BU}}})}{\max_j \exp({v^j_t * \tau_{\text{BU}}})} \\
    P(x'_k = n) &= 
    \begin{cases}
    p^k & \text{for } n=1 \\
    1 - p^k & \text{for } n=0
    \end{cases}
\end{align}

First the mean activation values for each feature map of VGG-16 are normalised to sum to 1. This normalisation accounts for differences in overall activation values between time steps and preserves relative differences between activation values. The activation values are then exponentiated and normalised by the maximum value across all feature maps. This ensures that the most active feature will receive a value of 1. Finally this is used as the probability that the $k^{\text{th}}$ entry of the particle state will be equal to one, as described by a Bernoulli distribution. In this way, a proportion of the particles are updated to represent the most active features given the current input. This is akin to bottom-up attention, whereby highly salient perceptual features capture ones attention in an involuntary manner. In our agent this serves to introduce a prior to attend to the highly active features of the current task. The free parameter $\tau_{\text{BU}}$ controls the strength of the bottom up attention, a higher value of $\tau_{\text{BU}}$ leads to a higher probability of the most active features being attended to and the least active features not being attended to.

In the observation step of a particle filter, particles are weighted based on the likelihood of the observed variable $O$ given the value of the latent variable $X$ represented by a particle. These weights are then used to re-sample the particles and update the posterior distribution ready for the next time step. In our approach, we introduce top-down attention during this step. We take our observed variable $O$ to be the return from a given state $R_t$. We calculate the likelihood of this return by using the particle state as the attention vector and calculating the error between the predicted state value and the return. Let $x^i$ denote the state of the $i^{\text{th}}$ particle, the likelihood for $x^i$ is calculated as follows:

\begin{align}
    A^i_k &= \frac{x^i_k}{\sum^K_{j=1} x^i_j} \label{eq:SA_norm} \\
    \delta^i &= (R_t - V(s_t; \mathbf{A^i}))^2 \label{eq:SA_error}\\
    P(R_t | x^i) &\propto \exp({-(\delta^i - \min_j \delta^j) * \tau_{TD}}) \label{eq:likelihood}
\end{align}

Where $R_t$ is the return from state $s_t$ and $V(s_t; \mathbf{A^i})$ is the predicted state value calculated by using the normalised particle state $x^i$ as the attention vector. The normalisation step in Equation \ref{eq:SA_norm} accounts for the different numbers of features that are attended to by different particles. Equation \ref{eq:SA_error} calculates the squared error between the return and predicted state value. The likelihood of the return $R_t$ is then proportional to this error value. This process can be seen as evaluating the accuracy of a particle's hypothesis over which features are relevant for the given task. If the particle's hypothesis is good then it will more accurately predict the target return and so will produce a larger likelihood. In this way we capture the effect of top-down attention, whereby a set of hypotheses are evaluated and the most accurate ones are considered for the next time step. $\tau_{TD}$ controls the strength of this top down attention; a larger value will more strongly penalise hypotheses that are inaccurate. 

Once the likelihoods have been calculated they are normalised to form a probability distribution and the particles are re-sampled with replacement:

\begin{align}
    P(x') &= \frac{\sum^N_{i=1} P(R_t | x') \mathbb{I}(x^i = x')}{\sum^N_{i=1}P(R_t | x^i)} \\
    x' &\sim P(x')
\end{align}

Once the re-sampling has been performed, the final step is to reset the value of the attention vector. This is done by setting the attention vector to be the mean of the particle states and then normalising the vector to sum to one:

\begin{align}
    \overline x_k &= \frac{1}{N} \sum^N_{n=1} x^n_k \\
    A_k &= \frac{\overline x_k}{\sum^K_{j=1} \overline x_j}
\end{align}

Where $N$ is the number of particles and $K$ is the number of features. The full algorithm used to update the attention vector can be seen in Algorithm \ref{alg:SA_particle_filter}.

\begin{algorithm}
	\caption{$\mathbf{UpdateAttention(v, s, R, \phi)}$}
	\begin{algorithmic}
	\State{Receive vector of normalised mean feature map values $\mathbf{v} = \{v_t, ..., v_{t+Z}\}$}
	\State{Receive vector of states $\mathbf{s} = \{s_t, ..., s_{t+Z}\}$}
	\State{Receive vector of returns $\mathbf{R} = \{R_t, ..., R_{t+Z}\}$}
	\State{Receive movement probability $\phi$}
	\State{}
	
    \State{\textbf{1. Perform Movement Step}}
    \State{Calculate feature probabilities}
    \State{$p^k = \dfrac{\frac{1}{Z} \sum^{t+Z}_{z=t} \exp({v^k_z * \tau_{\text{BU}}})} {\max_j \frac{1}{Z} \sum^{t+Z}_{z=t} \exp({v^j_z * \tau_{\text{BU}}})}$}
    \State{Update each particle state $x^i$ with probability $\phi$}
    \State{$P(x^i_k = n) = 
    \begin{cases}
    p^k & \text{for } n=1 \\
    1 - p^k & \text{for } n=0
    \end{cases} $}
	\State{}
	
	\State{\textbf{2. Perform Observe Step}}
	\State{Calculate the likelihoods of each particle $x^i$}
	\State{$A^i_k = \dfrac{x^i_k}{\sum^K_{j=1} x^i_j}$}
	\State{$\delta^i = \frac{1}{Z} \sum^{t+Z}_{z=t} (R_z - V(s_z; \mathbf{A^i}))^2$}
	\State{$P(\mathbf{R} | x^i) \propto \exp({-(\delta^i - \min_j \delta^j) * \tau_{TD}})$}
	\State{Re-sample particles based on calculated likelihoods}
	\State{$P(x') = \dfrac{\sum^N_{i=1} P(\mathbf{R} | x') \mathbb{I}(x^i = x')} {\sum^N_{i=1}P(\mathbf{R} | x^i)}$}
	\State{$x^i \sim P(x')$}
	\State{}
	
	\State{\textbf{3. Update Attention}}
	\State{Calculate mean particle state and normalise}
	\State{$\overline x_k = \frac{1}{N} \sum^N_{n=1} x^n_k$}
    \State{$A_k = \dfrac{\overline x_k} {\sum^K_{j=1} \overline x_j}$}
	\end{algorithmic}
	\label{alg:SA_particle_filter}
\end{algorithm}

\subsection{Deep Reinforcement Learning Algorithm}

The final component of our approach is a standard Deep Reinforcement Learning (RL) algorithm for selecting actions based on the results of the selective attention mechanism. Our approach can be used with a variety of Deep RL algorithms as long as they involve a value function. The use of a value function is critical because the particle filter uses the value predictions of each of its particles to calculate likelihood values for re-sampling (Equation \ref{eq:likelihood}). A value function therefore allows the particle filter to assess different hypotheses based on how predictive they are of reward. The specifics of each of the Deep RL algorithms used will be covered in the following sections based on the task being considered.

\section{Experiment 1 - Multiple Choice Task}

\subsection{Task}

\begin{figure}
	\centering
	\includegraphics[height=3cm, width=13.5cm]{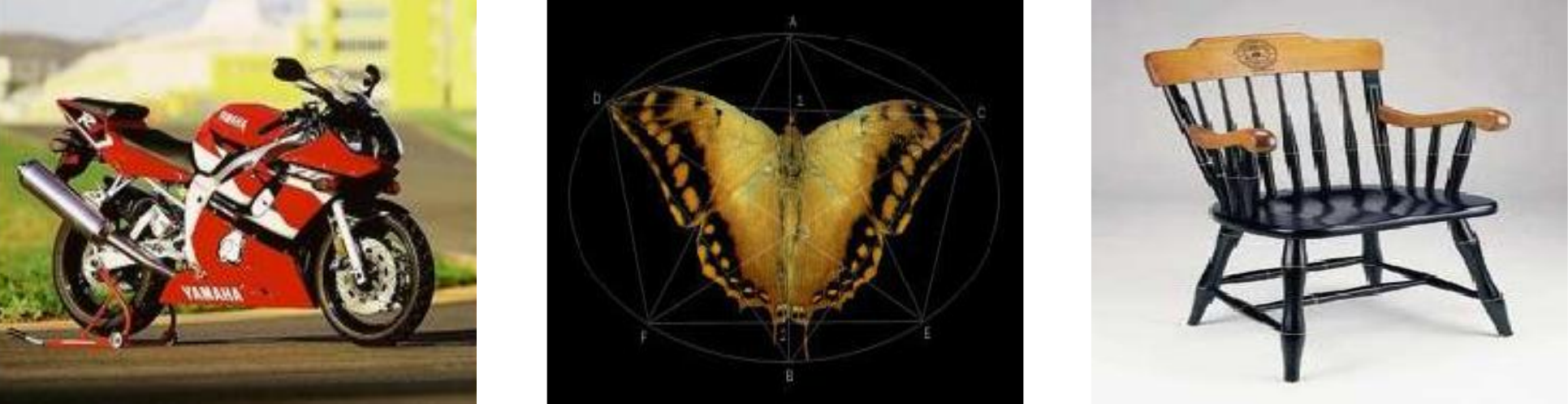}
	\caption{\textit{Example images from the Caltech 101 data set. Left image is from the 'motorbikes' category, the middle image is from the 'butterfly' category and the right image is from the 'chair' category. The agent was presented with three separate images, each from a randomly chosen category. In a given block of trials only one image category was associated with a reward of +1, the rest were associated with a reward of 0. The rewarded image category was chosen at random for each block of trials.}}
	\label{fig:Example_MC}
\end{figure}

The first task we explored consisted of a simple multiple choice format. We used the Caltech 101 data set \citep{fei2006one}, which consists of 101 object categories with approximately 40 to 800 images per category. Three categories from the Caltech 101 data set were chosen at random and the images from those categories were used for the multiple choice task (Figure \ref{fig:Example_MC}). The task consisted of 200 blocks of 50 trials. A single trial consisted of presenting the agent with 3 different natural images, one from each of the chosen categories. The images were presented separately, one after the other. For each block one of the categories was chosen at random and associated with a positive reward of +1 while the others were associated with no reward. The agent therefore has to work out which image is associated with a reward on each trial based on features of a specific category. Every time a new block starts the agent also has to adapt to the change in reward structure using only reward feedback.

To ensure that the agent did not learn to remember specific exemplars in the data set we had training and test phases. During the training phase the agent was able to update the parameters of the Deep RL network in response to reward feedback, however during the test phase it was not. With respect to attention, the agent was allowed to update its attention values during test but the network weights were kept fixed. The test phase used images from the chosen object categories that were not presented during training and consisted of 10 blocks of 50 trials.

\subsection{Reinforcement Learning Algorithm}

\begin{figure}
	\centering
	\includegraphics[height=6cm, width=4cm]{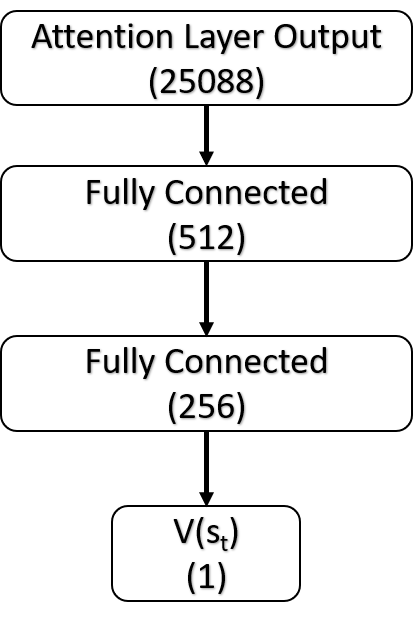}
	\caption{\textit{Architecture used in the multiple choice task for the deep reinforcement learning component of Selective Particle Attention ($SPA$). Numbers in brackets represent the number of units and each layer was fully connected. $V(s_t)$ corresponds to the value of a particular image.}}
	\label{fig:SA_MC_DeepRL}
\end{figure}

For the multiple choice task, the output of the attention layer was passed on to a simple 3 layer fully connected Deep Neural Network (DNN) (Figure \ref{fig:SA_MC_DeepRL}). The DNN was used to approximate the value of a given state, which in this case was the value of a given image. To select an action the value of each image was calculated and one of the images was chosen in an $\epsilon$-greedy manner, with $\epsilon=.2$. As each trial was based on a single choice the problem is equivalent to single step Markov Decision Process (MDP). We therefore trained the DNN after each trial to minimise the difference between the reward experienced and predicted value of the image that was chosen:

\begin{align}
    J(\theta) &= \tfrac{1}{2}(R_{t} - V(s_t; \theta, \mathbf{A}))^2 \\
    &= \tfrac{1}{2}(r_{t} - V(s_t; \theta, \mathbf{A}))^2 \\
    \nabla_{\theta} J(\theta) &= -(r_{t} - V(s_t; \theta, \mathbf{A})) \nabla_{\theta} V(s_t; \theta, \mathbf{A})
\end{align}

Where $\theta$ are the parameters of the DNN, $\mathbf{A}$ is the attention vector, $s_t$ is the image chosen at time $t$ and $r_{t}$ is the reward associated with the image chosen. We used RMSProp as an optimizer and the hyper-parameter values can be seen in Table \ref{table:MC_Parameters}. As mentioned in Section \ref{SA_methods} the weights of VGG-16 were kept constant and a particle filter was used to dynamically update the features that are being attended to. For the particular filter the return ($R_t$) in Equation \ref{eq:SA_error} was simply the reward experienced at the end of each trial ($r_{t}$). The full algorithm is shown in Algorithm \ref{alg:SA_MC}.

\begin{table}
	\caption{Hyper-parameter values used for the multiple choice task.}
	\label{table:MC_Parameters}
	\centering
	\begin{tabular}{lll}
		\toprule
		Parameter & Value & Description \\
		\midrule
		$N$ & 250 & Number of particles\\
		$\tau_{BU}$ & 10 & Strength of bottom-up attention \\
		$\tau_{TD}$ & 10 &  Strength of top-down attention\\
		$\epsilon$ & .2 &  Probability of selecting a random action\\
		$\lambda$ & .00025 & Learning rate for RMSProp\\
		$\kappa$ & .95 & Momentum for RMSProp\\
		$\iota$ & .01 & Constant for denominator in RMSProp\\
		\bottomrule
	\end{tabular}
\end{table}

\begin{algorithm}
	\caption{\textbf{Full algorithm for the multiple choice task}.}
	\begin{algorithmic}
	    \State{Initialize value function parameters with random weights $\theta$}
	    \State{Initialize attention vector $\mathbf{A} \gets \mathbf{1^K \cdot} \frac{1}{K}$}
	    \State{Initialize state of each particle $x^i \gets \mathbf{0}^K$}
	    \State{Initialise movement probability $\phi \gets 1$}
		\For{$t= 1, T$}
		    \State{Observe the three images $\{s^1_t, s^2_t, s^3_t\}$}
		    \State{With probability $\epsilon$ select a random action $a_t$}
		    \State{Otherwise $a_t \gets \arg\max_{a}(V(s^a_t); \theta, \mathbf{A})$}
		    \State{Receive reward $r_{t}$}
		    \State{Get mean feature maps $\overline{f}_t$ from random image $\sim U\{s^1_t, s^2_t, s^3_t\}$}
		    \State{Normalise feature maps $v^k_t \gets \overline{f}^k_t / \sum^K_{j=1}\overline{f}^j_t$}
		    \State{$UpdateAttention(\{v_{t}\}, \{s^{a_t}_{t}\}, \{r_{t}\}, \phi)$}
	        \State{$\theta \gets \theta + \alpha (r_{t} - V(s^{a_t}_{t}; \theta, \mathbf{A})) \nabla_{\theta} V(s^{a_t}_{t}; \theta, \mathbf{A})$}
	        \State{$\phi \gets .01$}
		\EndFor
	\end{algorithmic}
	\label{alg:SA_MC}
\end{algorithm}

\subsection{Results}

As a baseline to measure the effectiveness of our proposed attention mechanism we compared the performance of $SPA$ to a version of $SPA$ where each entry of the attention vector was set to a fixed value of $1/K$. This corresponds to attending to all features of VGG-16 and we refer to this approach as $SPA_{ALL}$. We also ran an ideal observer model on the multiple choice task to provide a measure of ceiling performance. The ideal observer model selects the image corresponding to the last rewarded object category. All models select a random action with probability $\epsilon$ in order to encourage exploration.

\begin{figure}
	\centering
	\includegraphics[height=6cm, width=13.5cm]{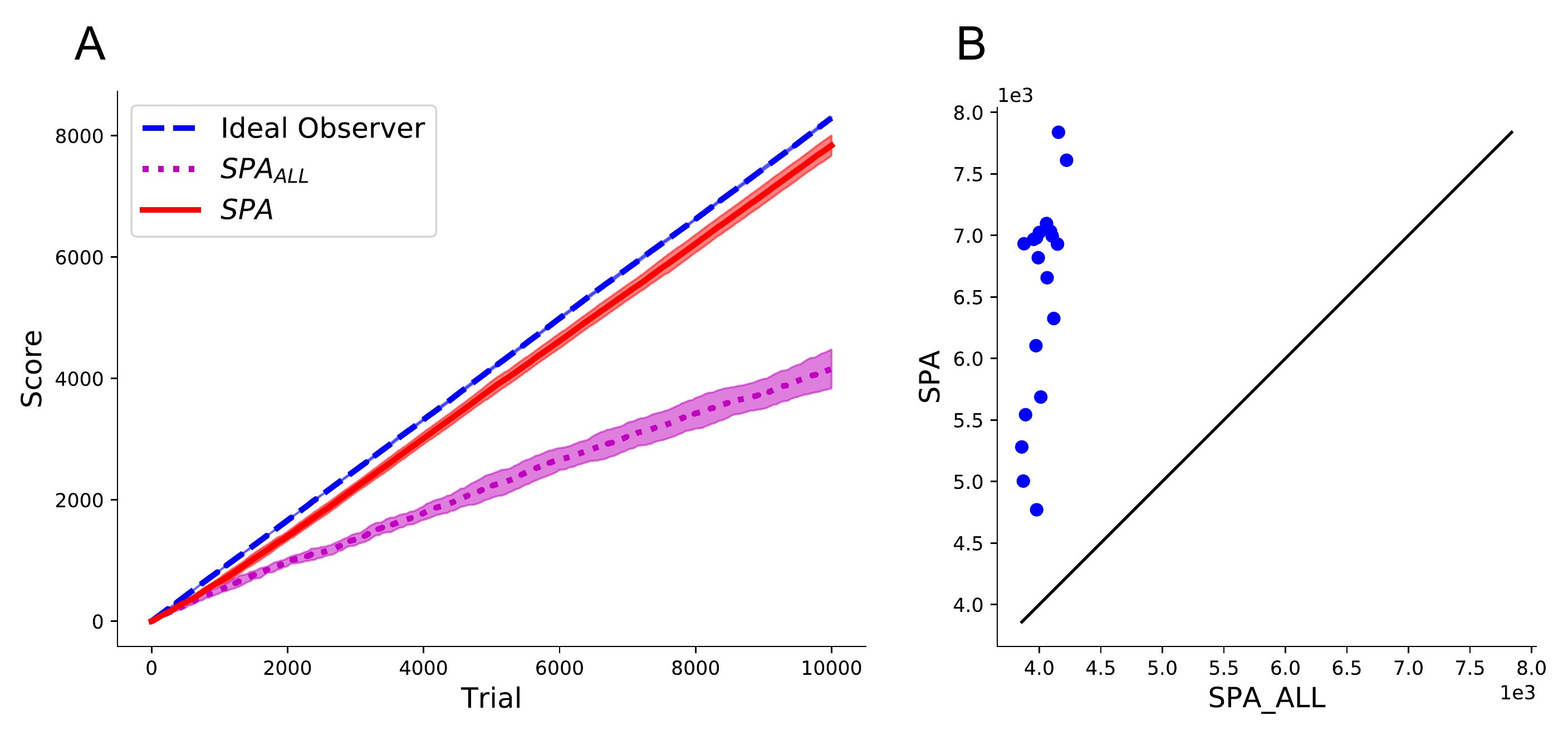}
	\caption{\textit{\textbf{(A)} Training scores over 5 random seeds for the leopards, faces and soccer balls image categories. The rewarded image category was changed randomly every 50 trials. Error bars represent one standard deviation. $SPA$ used selective attention to attend to features that it deemed useful for the current task. $SPA_{ALL}$ did not use selective attention and instead attended to every feature of VGG-16. The Ideal observer model represents ceiling performance. \textbf{(B)} Comparison of total training scores over 5 random seeds for 20 image category combinations between $SPA$ and $SPA_{ALL}$. Solid line represents equal performance between $SPA$ and $SPA_{ALL}$}}
	\label{fig:MC_Training}
\end{figure}

Figure \ref{fig:MC_Training}A shows the performance of $SPA$, $SPA_{ALL}$ and the ideal observer during training for one random combination of object categories over 5 random seeds. The rewarded image category was changed every 50 trials. In this example $SPA$ performs close to optimal as it shows a similar learning trajectory to the ideal observer. In comparison, $SPA_{ALL}$ performs poorly and is substantially worse than both $SPA$ and the ideal observer. To test the robustness of these findings we ran all three approaches over 5 random seeds on 20 different combinations of object categories. Figure \ref{fig:MC_Training}B shows the results of these simulations. $SPA$ out-performed $SPA_{ALL}$ during training for every combination of categories that we tested.

\begin{figure}
	\centering
	\includegraphics[height=6cm, width=13.5cm]{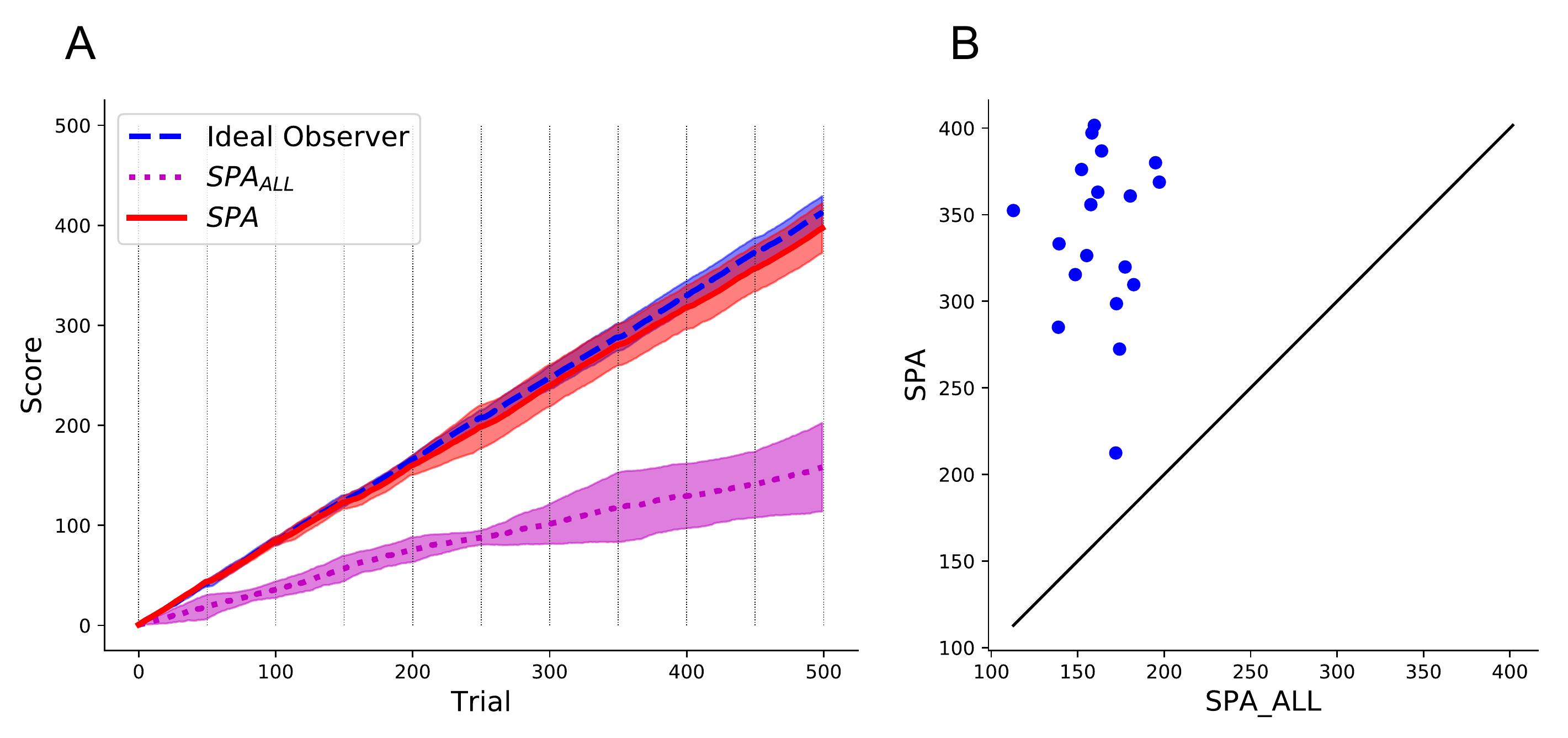}
	\caption{\textit{\textbf{(A)} Test scores over 5 random seeds for the leopards, faces and soccer balls image categories. The rewarded image category was changed randomly every 50 trials and is represented by a vertical black line. Error bars represent one standard deviation. \textbf{(B)} Comparison of total test scores over 5 random seeds for 20 image category combinations between $SPA$ and $SPA_{ALL}$. Solid line represents equal performance between $SPA$ and $SPA_{ALL}$}}
	\label{fig:MC_Test}
\end{figure}

While the dynamic attention mechanism of $SPA$ appeared to provide a substantial benefit during training, we also wanted to test whether this benefit generalized to unseen images. Figure \ref{fig:MC_Test}A shows the results of the three approaches on the test phase after the training seen in Figure \ref{fig:MC_Training}A. Again the rewarded image category was changed every 50 trials. Importantly the test blocks used images that were not used during training and all the weights of the Deep RL algorithm were frozen so that only the attention vector could change in the case of $SPA$. Again $SPA$ exhibited performance similar to that of the ideal observer, while $SPA_{ALL}$ showed significantly worse performance. Figure \ref{fig:MC_Test}B shows the test results over 5 random seeds for all 20 of the different category combinations. As with training, $SPA$ out-performed $SPA_{ALL}$ for all of the category combinations. These results suggests that the benefit of the attention mechanism of $SPA$ generalizes well to unseen images.

\begin{figure}
	\centering
	\includegraphics[height=7cm, width=9.5cm]{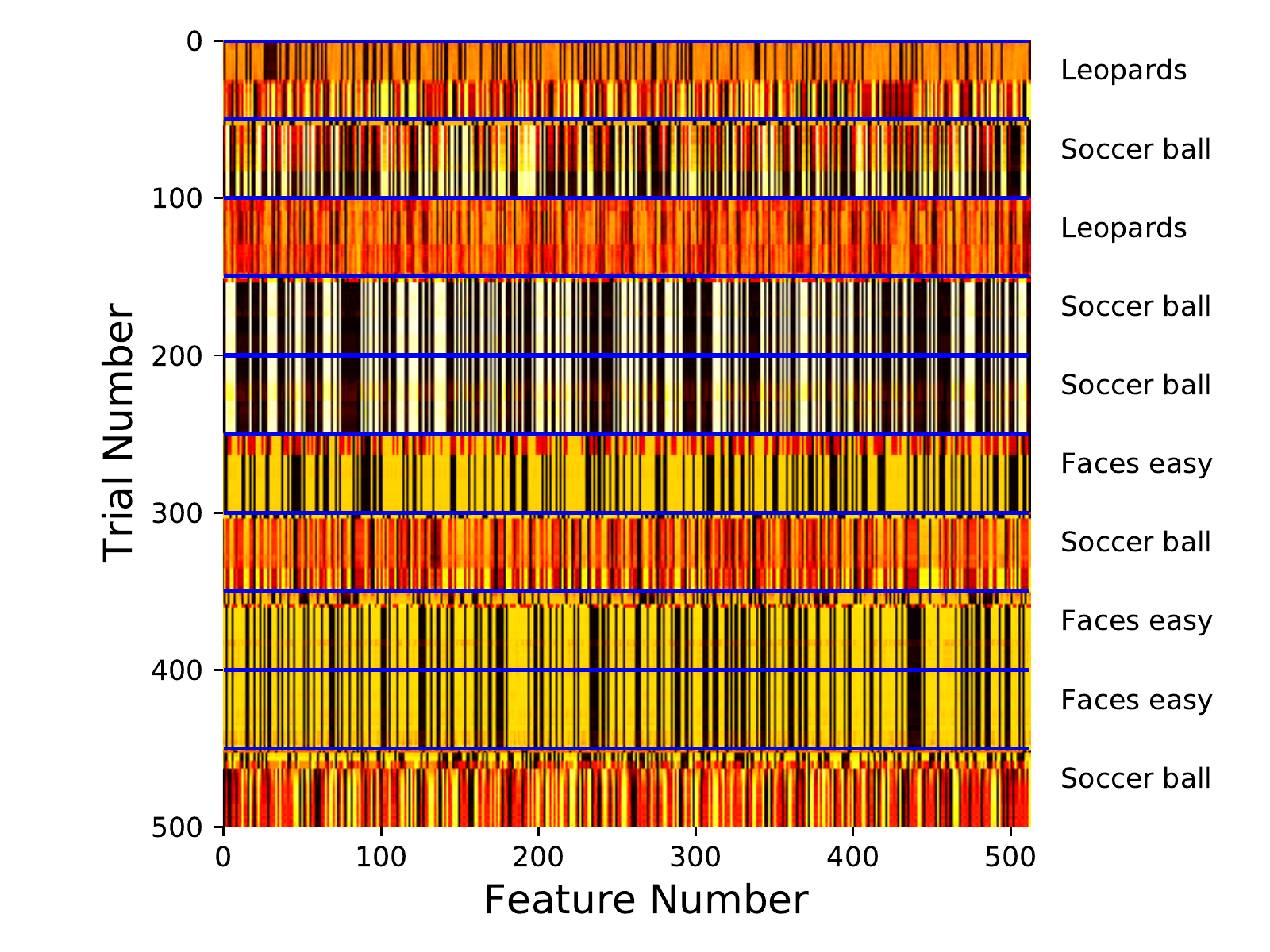}
	\caption{\textit{Example of the attention vector values for the leopards, faces and soccer balls image categories. Black represents a value of 0 and white represents a value of 1. The solid blue horizontal line represents a random change in the rewarded image category. The rewarded image category for a given block is presented on the right.}}
	\label{fig:MC_Test_Attention}
\end{figure}

Both the training and test results suggest that $SPA$ is able to cope with changes in the reward function by dynamically re-configuring existing representations for the purpose of state evaluation. Figure \ref{fig:MC_Test_Attention} shows an example of the attention vector during the test phase. The attention vector reliably changes when the target image category changes. This confirms that the attention mechanism of $SPA$ is able to use changes in the reward function to re-evaluate the features that need to be attended to. Interestingly, the attention vector is not the same every time a given image category is made the target. This is likely due to the fact that $SPA$ will be biased towards attending to features that are present in the first few images, which will be different for each block. In addition, there is likely to be a contextual effect of the rewarded image category in the previous block. For example, if in the previous block the category 'soccer ball' was rewarded and in the current block 'faces' are rewarded, then this might bias the selection of features that correspond to 'round' in the current block.

\begin{figure}
	\centering
	\includegraphics[height=7cm, width=13.5cm]{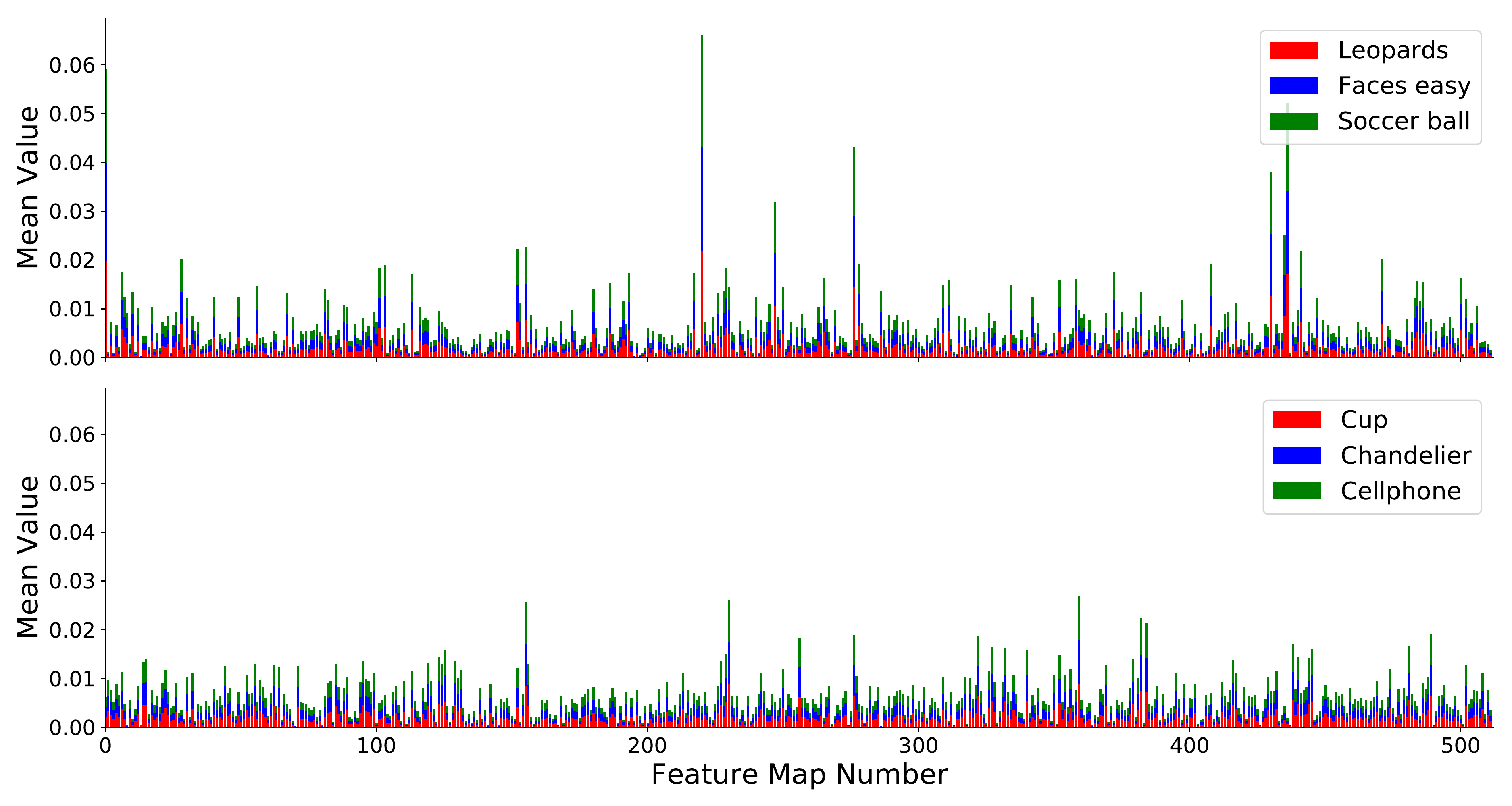}
	\caption{\textit{Average values for each of the feature maps of VGG-16. Top panel represents the values for the combination of image categories that lead to the best training performance of $SPA$. Bottom panel represents the values for the combination of image categories that lead to the worst training performance of $SPA$.}}
	\label{fig:Raw_Feature_Maps}
\end{figure}

The results shown in Figures \ref{fig:MC_Training}B and \ref{fig:MC_Test}B indicate that the performance of $SPA$ can vary depending on the combination of image categories that are chosen. This suggests that $SPA$ finds it easier to discriminate between certain image categories compared to others based on their features. Figure \ref{fig:Raw_Feature_Maps} shows the mean feature map values of VGG-16 for the image categories that $SPA$ performed best (leopards, faces and soccer balls) and worst (cups, chandeliers and cellphones) on. In the best case scenario, the feature maps contain several features that are substantially more active than others. This likely provides a good substrate for bottom-up attention because there are a handful of features that are reliably more active than the others, which corresponds to a strong prior over hypotheses. In comparison, in the worse case scenario, the features take on a more uniform distribution of activation values and so the prior over hypotheses is weaker.

\begin{figure}
	\centering
	\includegraphics[height=5cm, width=13.5cm]{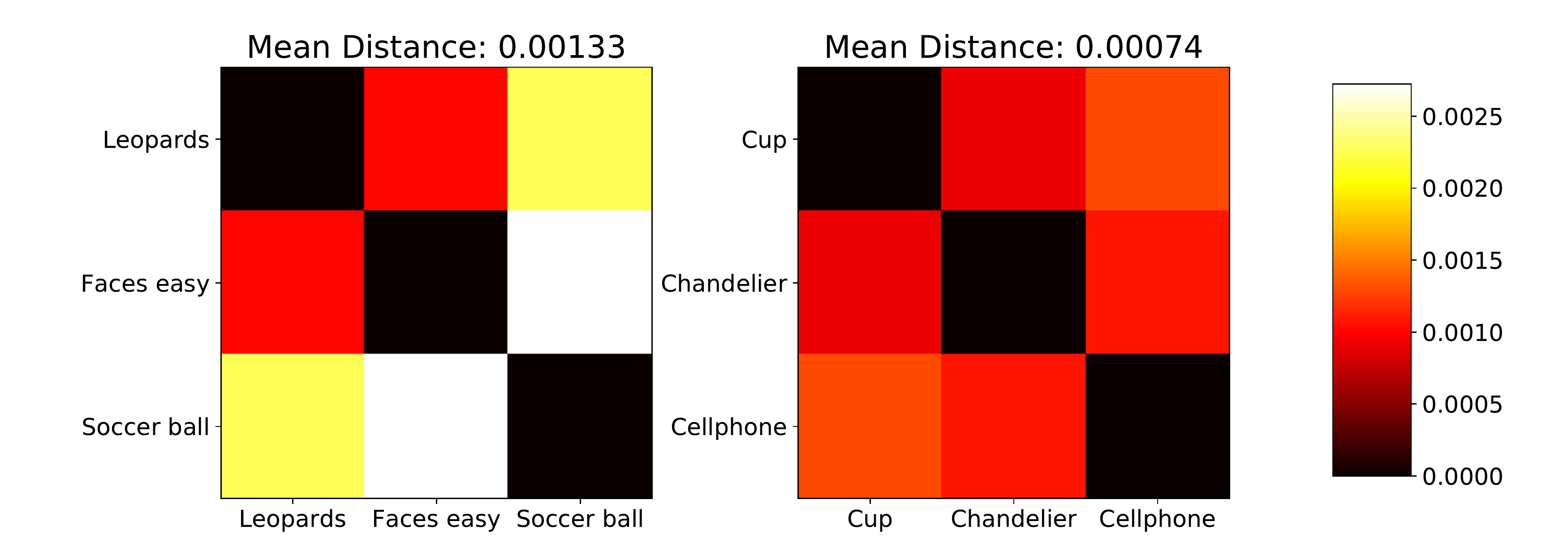}
	\caption{\textit{Euclidean distances between the average feature map values of VGG-16. The left panel shows the distances between the image categories that lead to the best training performance of $SPA$. The right panel shows the distances between the image categories that lead to the worst training performance of $SPA$.}}
	\label{fig:L2_Norms}
\end{figure}

Bottom-up attention aside, having a few highly active features is only useful for the multiple choice task if they help to discriminate between the different image categories. As seen in Figure \ref{fig:Raw_Feature_Maps}, each image category produces 512 average feature values, which can then be expressed as a vector in Euclidean space. Figure \ref{fig:L2_Norms} shows the Euclidean distance between these vectors for both the best and worst case scenarios. In the best case scenario, the euclidean distance between the majority of the image categories is larger than in the worst case scenario. In addition the mean euclidean distance over all pairwise comparisons in the best case scenario is nearly double that of the worst case scenario. This suggests that the categories in the best case scenario are easier to discriminate between because they are further apart in euclidean space. This is likely to help the top-down attention of $SPA$ because each particle will produce very different value estimates depending on the image category being attended to.

\begin{comment}
- visualize features

- remove bottom-up component

- remove top-down component

- is it the combination of both that are important?
\end{comment}

\section{Experiment 2 - Object Collection Game}

\subsection{Task}

\begin{figure}
	\centering
	\includegraphics[height=4cm, width=13.5cm]{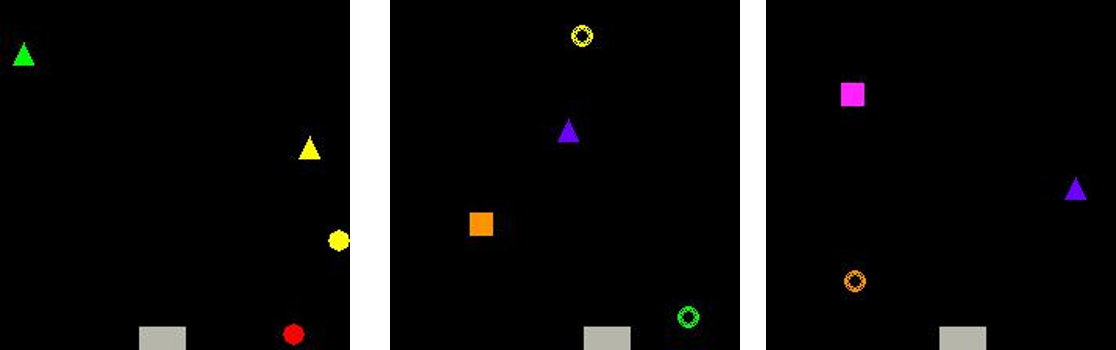}
	\caption{\textit{Example screenshots of the object collection game. The agent is in control of the grey rectangle and can move it either left or right to catch shapes that move from the top of the screen to the bottom. Only catching a specific shape leads to a reward of +1 and the target shape is chosen at random.}}
	\label{fig:SA_OC}
\end{figure}

Having demonstrated the effectiveness of $SPA$ in a single step Markov Decision Process (MDP), we next wanted to investigate whether it could handle multi-step MDPs from raw pixel input. To this end we designed an object collection game, which consisted of a simple 2D video game that was made using Pygame (www.pygame.org). The agent controls a block at the bottom of the screen and can move it either left or right at each time step. A random object that can vary in shape and colour is randomly generated every 60 seconds at the top of the screen and moves downwards to the bottom of the screen. A variety of screen shots of the game can be seen in Figure \ref{fig:SA_OC}. Each time the game is run a random shape is chosen to be the target shape. The goal of the agent is to collect the target shape when it reaches the bottom of the screen, at which point it receives a reward of 1. Collisions with any of the other shapes result in a reward of 0. This task is designed to explore the ability of an agent to focus on particular features of the environment i.e. attend to the target shape and ignore the other shapes and colours. Each trial lasted 60 seconds after which a new trial would begin with the same target shape.

\subsection{Reinforcement Learning Algorithm}

\begin{figure}
	\centering
	\includegraphics[height=8cm, width=4.5cm]{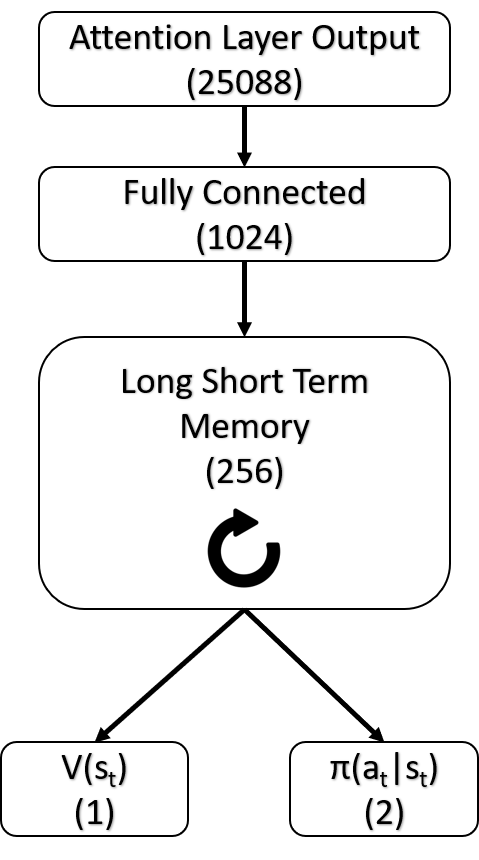}
	\caption{\textit{Architecture used in the object collection game for the deep reinforcement learning component of Selective Particle Attention ($SPA$). Numbers in brackets represent the number of units and each layer was fully connected. An actor-critic architecture was used so that the network output both a state value and a probabilistic policy in the form of a soft-max distribution.}}
	\label{fig:SA_OC_DeepRL}
\end{figure}

The only modification to the architecture of SPA from the previous experiments was a change in the Deep Reinforcement Learning (RL) algorithm. For the Deep RL algorithm we chose to use an actor-critic architecture. We chose this setup to explore whether $SPA$ could work when the network also had to compute a policy. The agent received every 8th frame as input. The output of the attention layer was passed to a fully connected layer followed by a Long-Short Term Memory (LSTM) network. The output of the LSTM was then passed onto two final layers, one that represented a softmax distribution over actions and another that represented the value of the current state (Figure \ref{fig:SA_OC_DeepRL}). The network therefore resembled an advantage actor critic (A2C) architecture \citep{mnih2016asynchronous}, whereby the predicted state value from the critic was used to calculate the advantage function. This advantage function was then used to train the actor. The cost function of the critic was the mean squared error between the return and the predicted value:

\begin{align}
    J(\theta) &= \tfrac{1}{2}(R_t - V(s_t; \theta, \mathbf{A}))^2 \label{eq:SA_critic}\\
    \nabla_{\theta} J(\theta) &= -(R_t - V(s_t; \theta, \mathbf{A})) \nabla_{\theta} V(s_t; \theta, \mathbf{A})
\end{align}

Where $\theta$ are the parameters of the DNN, $\mathbf{A}$ is the attention vector, $s_t$ is the frame of the game at time $t$ and $R_t$ is the return from state $s_t$. The actor was updated using the advantage calculated from the critic:

\begin{align}
    A(s_t, a_t) &= Q(s_t, a_t) - V(s_t) \\
    &= R_t - V(s_t; \theta, \mathbf{A}) \label{eq:SA_advantage}\\
    \nabla_{\theta} L(\theta) &= A(s_t, a_t) \nabla_{\theta} \ln{\pi(a_t \mid s_t ; \theta, \mathbf{A})} + \beta \nabla_{\theta} H(\pi(a_t \mid s_t ; \theta, \mathbf{A}))
\end{align}

Where $H$ represents the entropy of the softmax distribution over actions and is included to encourage exploration. $\beta$ is a free parameter that controls the strength of this entropy term and was set to 0.01 for all simulations. In all cases, the return $R_t$ from state $s_t$ was estimated using n-step Temporal Difference (TD) learning:

\begin{align}
    R^n_t &= r_{t} + \gamma r_{t+1} + ... \gamma^{n - 1} r_{t+n-1} + \gamma^n V(s_{t+n}; \theta, \mathbf{A}) \\
    &= \sum^{n-1}_{i=0} \gamma^i r_{t+i} + \gamma^n V(s_{t+n}; \theta, \mathbf{A})
\end{align}

This was used for calculating the target for the critic via Equation \ref{eq:SA_critic} and for calculating the advantage in Equation \ref{eq:SA_advantage}. The training procedure was the same as in \citet{mnih2016asynchronous} with $t_{max}$ set to 10. More specifically, the agent selects actions up until $t_{max}$ or the trial ends. At this point the the gradients are calculated for each of the n-step TD-learning updates using the longest possible n-step return. For example, the update for the last state will be a one-step update whereas the update for the first state will be a $t_{max}$-step update. These gradients are all applied in a single update.

The calculated returns using the aforementioned procedure were also used in Equation \ref{eq:SA_error} for calculating the observation step of the particle filter. However, the movement and update steps of the particle filter were not performed on every training step (when $t_{max}$ or the trial ends). Instead they were performed every $c_{max}$ training steps, we found that this improved both stability and reduced training time. The full algorithm is shown in Algorithm \ref{alg:SA_A2C} and Table \ref{table:OC_Parameters} shows the hyper-parameter values used.

\begin{table}
	\caption{Hyper-parameter values used for the object collection game.}
	\label{table:OC_Parameters}
	\centering
	\begin{tabular}{lll}
		\toprule
		Parameter & Value & Description \\
		\midrule
		$N$ & 250 & Number of particles\\
		$\tau_{BU}$ & 1 & Strength of bottom-up attention\\
		$\tau_{TD}$ & 10 & Strength of top-down attention\\
		$c_{max}$ & 1000 & Frequency of attention updates\\
		$t_{max}$ & 10 & Maximum length of return\\
		$\beta$ & 0.01 & Exploration strength\\
		$\gamma$ & .99 & Discount factor for future rewards\\
		$m$ & 8 & Number of frames skipped\\
		$\alpha$ & 0.0001 & Learning rate for Adam optimizer\\
		\bottomrule
	\end{tabular}
\end{table}

\begin{algorithm}
	\caption{\textbf{Full algorithm for the object collection game}.}
	\begin{algorithmic}
	    \State{Initialize A2C network parameters with random weights $\theta$}
	    \State{Initialize time step and attention step counters $t \gets 0$, $c \gets 0$}
	    \State{Initialize attention vector $\mathbf{A} \gets \mathbf{1^K} \cdot \frac{1}{K}$}
	    \State{Initialize state of each particle $x^i \gets \mathbf{0}^K$}
	    \State{Initialise movement probability $\phi \gets 1$}
		\For{$e= 1, E$}
		    \State{$d\theta_{\pi}, d\theta_v \gets 0$}
    		\State{$t_{start} \gets t$}
    		\State{Get state $s_t$ and mean feature maps $\overline{f}_t$}
    		\Repeat{}
    		    \State{Normalise mean feature maps $v^k_t \gets \overline{f}^k_t / \sum^K_{j=1}\overline{f}^j_t$}
        		\State{Perform action $a_t$ using policy $\pi(a_t \mid s_t; \theta_{\pi}, \mathbf{A})$}
        		\State{Receive reward $r_{t}$, state $s_{t+1}$ and mean feature maps $\overline{f}_{t+1}$}
        		\State{$t \gets t + 1$}
        	\Until{terminal $s_t$ \textbf{or} $t - t_{start} == t_{max}$}
    	    \State{$R_t =
    			\begin{cases} 
    			0 & \text{if terminal } s_t \\
    			V(s_t; \theta_v, \mathbf{A}) & \text{otherwise}
    			\end{cases}$}
    		\For{$i \in \{t-1, ..., t_{start}\}$}
    		    \State{$R_i \gets r_i + \gamma R_{i+1}$}
    		\EndFor
    		
    		\State{$c \gets c + 1$}
		    \If{$c == c_{max}$}
		        \State{$\mathbf{v} \gets \{v_{t_{start}}, ..., v_{t-1}\}$}
		        \State{$\mathbf{s} \gets \{s_{t_{start}}, ..., s_{t-1}\}$}
		        \State{$\mathbf{R} \gets \{R_{t_{start}}, ..., R_{t-1}\}$}
    		    \State{$UpdateAttention(\mathbf{v}, \mathbf{s}, \mathbf{R}, \phi)$}
    		    \State{$\phi \gets .01$}
    		    \State{$c \gets 0$}
		    \EndIf
		    
		    \For{$R \in \{R_{t_{start}}, ..., R_{t-1}\}$}
    		    \State{$d\theta_{\pi} \gets d\theta_{\pi} + \nabla_{\theta_{\pi}} \log \pi(a_t \mid s_t; \theta_{\pi}, \mathbf{A})(R - V(s_t; \theta_v, \mathbf{A}))$}
    		    \State{$d\theta_v \gets d\theta_v + (R - V(s_t; \theta_v, \mathbf{A}))\nabla_{\theta_v}V(s_t; \theta_v, \mathbf{A})$}
    		\EndFor
    		
    		\State{$\theta_{\pi} \gets \theta_{\pi} + \alpha d\theta_{\pi}$}
    		\State{$\theta_v \gets \theta_v + \alpha d\theta_v$}
		\EndFor
	\end{algorithmic}
	\label{alg:SA_A2C}
\end{algorithm}

\subsection{Results}

As with the multiple choice task, we compared $SPA$ to a baseline approach that attended to all features of VGG-16, which we refer to as $SPA_{ALL}$. In addition to $SPA_{ALL}$, we also included a condition that set the attention vector to a random binary vector, which was then normalised to sum to 1. This condition was included to account for the fact that random feature reduction may lead to improved performance and we refer to it as $SPA_{RANDOM}$. All models were ran on the object collection game for a total of 1000 episodes over 5 random seeds.

\begin{figure}
	\centering
	\includegraphics[height=9cm, width=13.5cm]{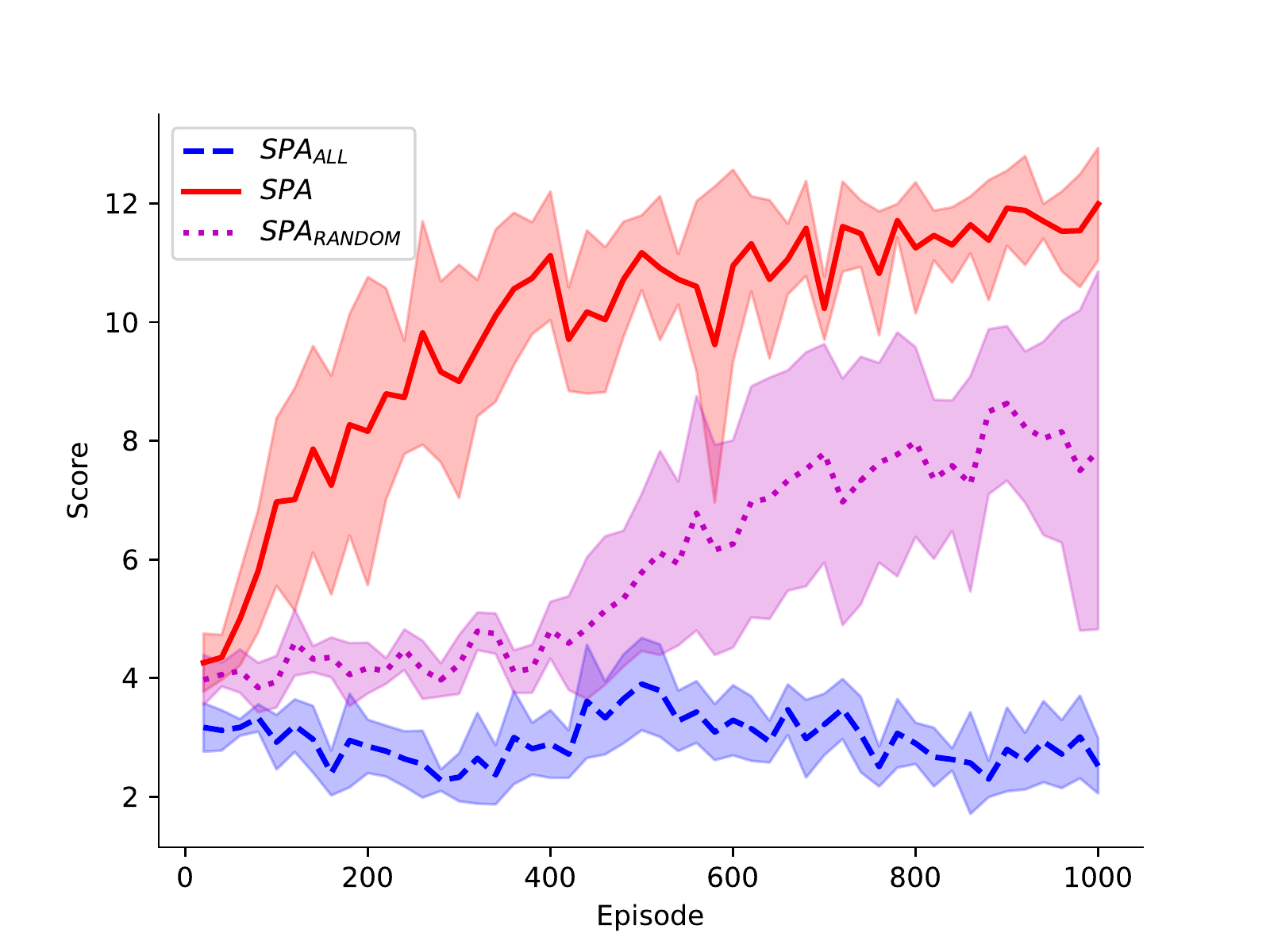}
	\caption{\textit{Performance of $SPA$, $SPA_{ALL}$ and $SPA_{RANDOM}$ on the object collection game. Each point is the average score over the last 20 episodes. Error bars represent one standard deviation. Each approach was run over 5 different random seeds and the error bars represent one standard deviation. $SPA$ used selective attention to attend to features that it deemed useful for the current task. $SPA_{ALL}$ did not use selective attention and instead attended to every feature of VGG-16. $SPA_{RANDOM}$ attended to a random subset of features, which changed with each random seed.}}
	\label{fig:OC_FirstLevel}
\end{figure}

Figure \ref{fig:OC_FirstLevel} shows the results of $SPA$, $SPA_{ALL}$ and $SPA_{RANDOM}$ on the object collection game. $SPA$ significantly out-performed $SPA_{ALL}$ and $SPA_{RANDOM}$ over the course of learning. $SPA_{ALL}$ saw the worst performance showing no evidence of learning over the 1000 episodes. This suggests that naively learning over all features is a highly ineffective strategy given a limited amount of experience. In contrast, $SPA_{RANDOM}$ showed evidence of learning after around 500 episodes. This learning was highly variable as would be expected given that the features were randomly selected for on each random seed. Nevertheless this demonstrates that simply reducing the number of features at random is sufficient to provide a learning benefit. Out of all of the approaches, $SPA$ performed the best for several reasons. Firstly, $SPA$ began improving its score almost immediately after the first couple of episodes. This onset of learning is noticeably earlier than the other approaches. Not only did learning occur earlier but it was also much faster, as indicated by the sharper increase in score compared to the other approaches. As a result, by the end of learning $SPA$ was able to achieve a better score on the final trial compared to the other approaches. Importantly, $SPA$ was also more robust than $SPA_{RANDOM}$ as indicated by the smaller standard deviation in Figure \ref{fig:OC_FirstLevel}. Overall, these results suggest that the particle filter mechanism of $SPA$ is able to identify features useful for value computation and that this translates into a substantial learning benefit given the current task.

\begin{figure}
	\centering
	\includegraphics[height=9cm, width=13.5cm]{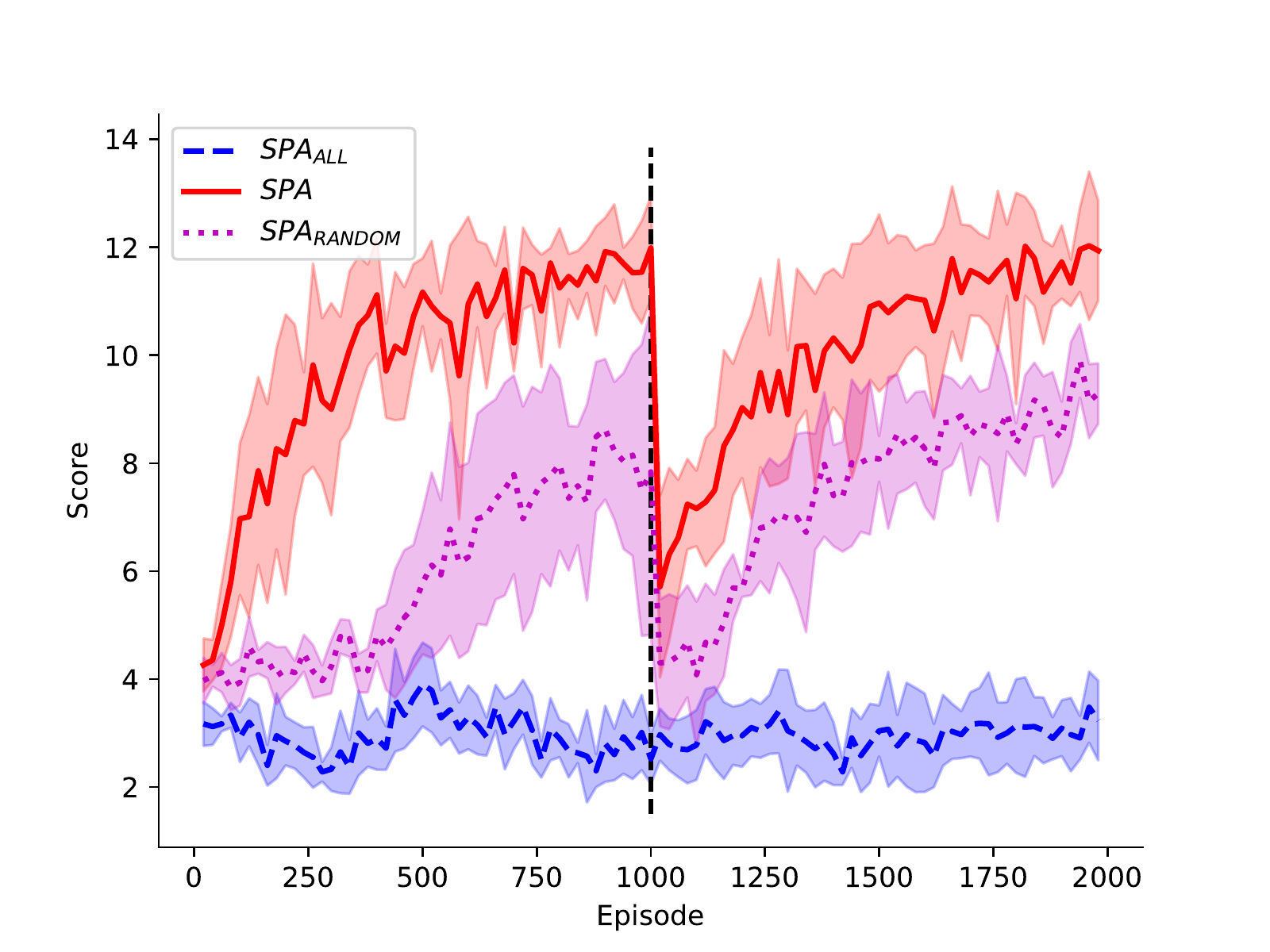}
	\caption{\textit{Performance of $SPA$, $SPA_{ALL}$ and $SPA_{RANDOM}$ on two sequential levels of the object collection game. The dashed line represents when the rewarded object was changed without any prior warning. All other details are the same as Figure \ref{fig:OC_FirstLevel}.}}
	\label{fig:OC_AllLevels}
\end{figure}

One benefit of an effective selective attention mechanism is that attention can be altered in response to changing task requirements. This allows for increased behavioural flexibility as learning can be rapidly guided to different sets of features. To investigate whether the attention mechanism of $SPA$ conferred increased flexibility we trained the models on 1000 episodes of the object collection game and then randomly changed the rewarded object for the next 1000 episodes. This is akin to changing the reward function of the environment and requires the agent to re-evaluate its policy. Figure \ref{fig:OC_AllLevels} shows the results of $SPA$, $SPA_{ALL}$ and $SPA_{RANDOM}$ on the change in reward function. As before, $SPA_{ALL}$ demonstrated no evidence of learning on the first 1000 episodes and this persisted for the second 1000 episodes. In comparison, $SPA_{RANDOM}$ showed a marked decrease in performance back to baseline levels as soon as the reward function changed and was slow to recover performance. $SPA$ also showed a marked decrease in performance immediately after the change in reward function. This is to be expected as the agent had no prior warning that the reward function was about to change. However, unlike $SPA_{RANDOM}$, the performance of $SPA$ dropped to above basline levels and remained significantly higher than $SPA_{RANDOM}$ throughout the recovery in performance. These results suggest that the particle filter mechanism of $SPA$ is better equipped to handle changes in the environment by quickly re-evaluating which features are important for the current task.

Randomly changing the rewarded object corresponds to a change in the reward function of the task. However, other aspects of the RL problem can change. Most notably the state space of the task can change, which corresponds to a change in the perceptual input of the agent. We therefore wanted to evaluate how well $SPA$ could deal with changes in the state space. To test this we first trained all three approaches on 1000 episodes of a version of the object collection game where all the objects were the same shape and colour. The agent was given a reward of +1 for obtaining these objects. Upon completing 1000 episodes, the shape and colour of the objects were changed to a different shape and colour, and the agent was trained for a further 1000 episodes. This regime corresponds to a change in state space because the objects present in the first 1000 episodes are different to the ones present in the second 1000 episodes.

\begin{figure}
	\centering
	\includegraphics[height=9cm, width=13.5cm]{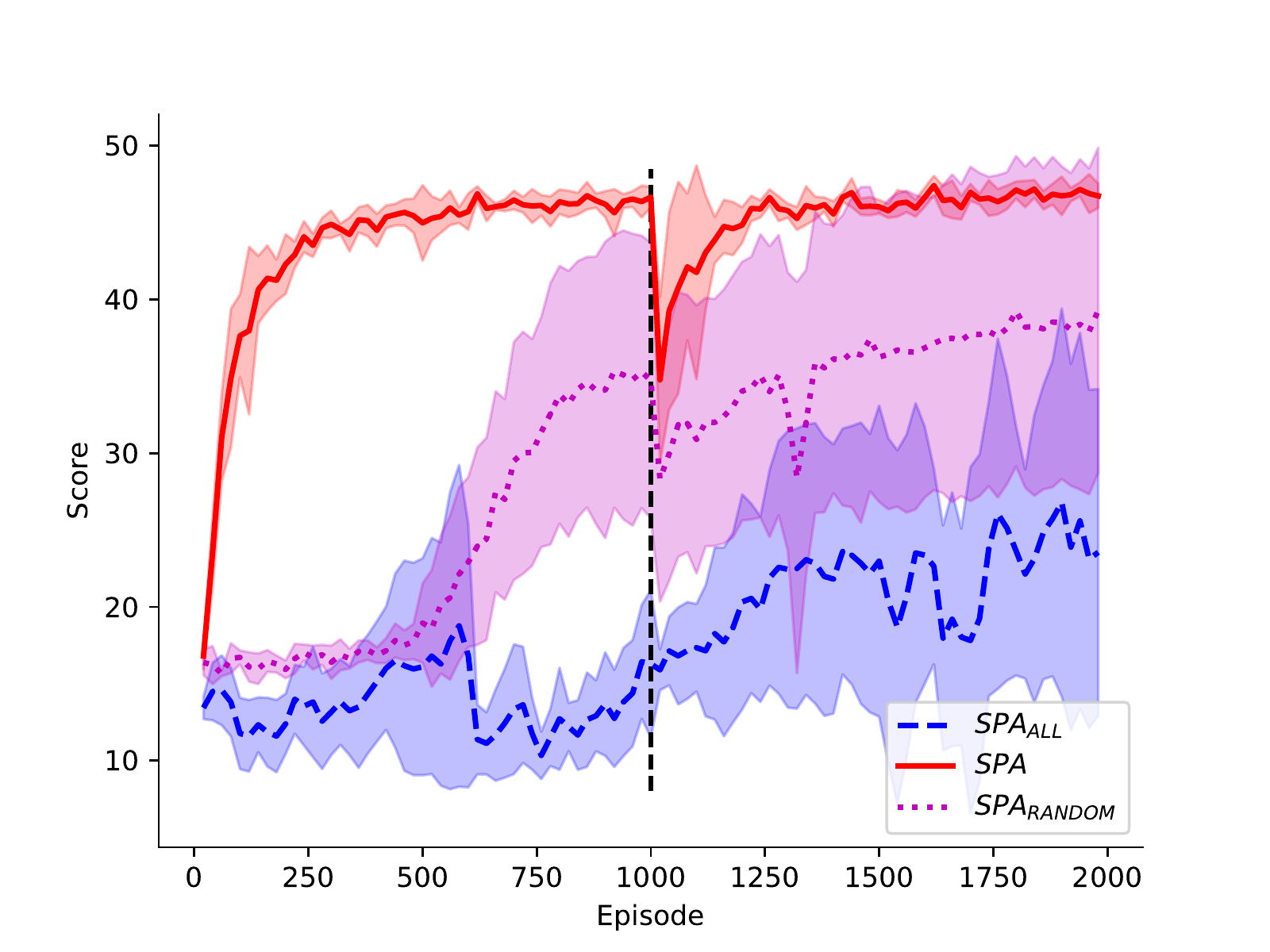}
	\caption{\textit{Performance of $SPA$, $SPA_{ALL}$ and $SPA_{RANDOM}$ on two sequential levels of the object collection game where all shapes were the same colour and shape. The dashed line represents when the shape and colour of these shapes changed without any prior warning. All other details are the same as Figure \ref{fig:OC_FirstLevel}.}}
	\label{fig:OC_AllLevelsState}
\end{figure}

Figure \ref{fig:OC_AllLevelsState} shows the results of the three approaches on the change in state space task. As before, for the first 1000 episodes $SPA$ demonstrated evidence of learning that occurred earlier, faster and more robustly than the other approaches. Upon the change in state space, $SPA$ showed a slight decrease in performance but was quickly able to recover from it. In comparison, $SPA_{RANDOM}$ also saw a drop in performance but the rate of recovery was slower, while $SPA_{ALL}$ continued its slow gradual learning. For both $SPA_{RANDOM}$ and $SPA_{ALL}$, performance for the second 1000 episodes was highly variable and significantly lower than $SPA$. These results further support the idea that the selective attention mechanism of $SPA$ is able to quickly adapt to changes in the current task by re-orientating its attention towards a different set of features. In addition, the consistently high performance of $SPA$ and its fast rate of recovery during the change in state space indicates that much of the knowledge learnt by the Deep RL algorithm in the first 1000 episodes was still of use in the second 100 episodes. This suggests that the selective attention mechanism of $SPA$ is also able to promote the transfer of knowledge between tasks.

\section{Discussion}

When presented with a visual scene we need to be able to quickly identify the relevant features based on our current goal. For example, if we are in a forest and are looking to start a fire then we need to attend to features indicative of dry wood. Conversely, if we are thirsty then we must attend to features that are indicative of a water source. This goal driven modulation of perceptual features is often referred to as selective attention and is thought to be the responsibility of the Pre-Frontal Cortex (PFC). Deep Reinforcement Learning (RL) represents a promising avenue for modelling how humans map raw perceptual input to goal-driven behaviour. Deep RL systems typically rely on the incremental learning of representations via backpropagation in Deep Neural Networks (DNNs). As a result they lack the ability to quickly adjust their representations given a sudden change in task. 

To address this issue and imbue Deep RL agents with the ability to perform selective attention we proposed a novel algorithm called \textit{Selective Particle Attention ($SPA$)}. The premise of $SPA$ is that a particle filter can be used to dynamically attend to subsets of features based on the current goal. During the movement step of the particle filter we incorporate bottom-up attention; particles are biased to attend to features that are highly active. During the observation step we incorporate top-down attention; particles that are better at predicting reward are more likely to be re-sampled. By iterating over these two steps the particle filter is able to perform a form of selective attention that combines bottom-up and top-down attention. From a conceptual point of view each particle can be thought of as a hypothesis about which sets of features are important for the task at hand. These hypotheses are biased towards features that are highly active and are constantly evaluated based on their ability to predict reward.

We evaluated $SPA$ on two different tasks. The first task was a multiple choice task involving naturalistic images. Three object categories were chosen at random and on each trial the agent was presented with an image from each of the categories. Trials were organised into blocks and for any given block only one of the object categories would result in a reward. $SPA$ was able to achieve close to ceiling performance on this task for several examples and dramatically out-performed a naive version that attended to all features. Inspection of $SPA$'s attention vector showed that it was able to quickly change its configuration in response to changes in the reward function. This highlights how selective attention can be used to quickly respond to unannounced changes in the environment and improve the efficiency of learning.

\begin{comment}
 the bottom-up component was an integral part of the performance benefit seen in the multiple choice task (XXX).
\end{comment}

The power of RL is its ability to deal with temporal dependencies and to make actions that lead to reward in the future. The aforementioned multiple choice task can be viewed as a single step Markov Decision Process (MDP). We therefore wanted to test if $SPA$ could be applied to more complex domains and multi-step MDPs. With this in mind we chose the second task to be a simple 2D video game. The agent was in control of a rectangular block that could be moved left and right. The goal of the agent was to collect objects that moved from the top of the screen to the bottom based on their shape. As with the multiple choice task, $SPA$ was able to perform significantly better than a naive approach that attended to all features. It also performed better than an approach that selected a random subset of features to attend to. This provides evidence that $SPA$ can improve the efficiency of learning on complex RL problems that require temporal dependencies. In addition, it demonstrates that $SPA$ is not dependent on a specific Deep RL algorithm as long as the algorithm uses a form of value approximation.

Interestingly, in order to successfully apply $SPA$ to the 2D object collection game the attention vector was only updated every 1000 time-steps rather than on every time step. This was necessary because it allowed the Deep RL algorithm time to respond to the change in attention and attempt to learn a useful function of the currently attended features. This is akin to a person learning based on a single hypothesis for a fixed amount of time before deciding whether to change to another hypothesis or not. This may explain why attentional inertia is prevalent in children and adults \citep{anderson1987attentional, burns1993attentional, richards2004attentional, longman2014attentional}, as the brain requires time to evaluate a given hypothesis before deciding whether to switch attention.  Future work should systematically explore this apparent trade-off between the potential benefits of switching to a new hypothesis and the time needed to sufficiently evaluate a hypothesis. 

Part of the reason for the design of the object collection game was that it allowed for the rewarded shape to be easily changed during learning. This change can be manipulated to correspond to a change in the reward function or state space. We found that $SPA$ was significantly better equipped to deal with such changes compared to other naive approaches that either attended to all or a random subset of features. In particular, $SPA$ showed an extremely fast recovery in response to a change in state space. This is often considered an extremely hard problem as deep neural networks typically fail catastrophically when the input distribution is changed \citep{lake2017building}. Nevertheless, the small drop in performance and rapid recovery to asymptotic levels suggests that $SPA$ was able to transfer much of the knowledge that it had acquired before the change in state space. These results further support the findings of the multiple choice task, which demonstrated $SPA$'s improved ability to deal with changes in the environment.

We propose that the ability of $SPA$ to focus on a subset of features based on the current task is beneficial for several reasons. Firstly, it helps to reduce the dimensionality of the problem, which reduces the impact of noisy features and the complexity of the function that needs to be learnt. This improves generalization because the learnt function does not fit spurious features and therefore ignores any changes to them. This benefit of dimensionality reduction can even be seen in the case of $SPA_{RANDOM}$, which out-performed $SPA_{ALL}$ on the object collection game. $SPA$ takes this one step further however, by providing targeted dimensionality reduction rather than selecting a random subset of features. Furthermore, the selective attention of $SPA$ helps to reduce interference by guiding learning onto different sets of features based on the task at hand. This results in the learning of different sub-networks that can be quickly identified based on the current task. This greatly improves the capacity of the Deep RL algorithm to represent different functions and to switch between them in the face of changes to the environment.

Selective attention in $SPA$ is implemented using a particle filter that captures the influences of both bottom-up and top-down attention. Previous work has already shown that particle filters may represent a viable computational account of selective attention during learning based on their ability to fit eye-tracking data \citep{radulescuparticle}. Here we extend this work to show how they can be interfaced with Deep RL principles in order to learn which features to attend to just from raw pixel values and external reward signals. One of the primary benefits of using a particle filter to implement selective attention is that it relies on sampling to produce an approximate value of a hidden variable. This is important because the potential number of feature combinations that need to be evaluated for a given task can be extremely large. In our case there were $2^{512}$ potential feature combinations and so it would be computationally infeasible to evaluate all of them. However, by using only 250 particles we were still able to converge to a satisfactory solution thanks to the iterative re-sampling procedure of the particle filter. Future work should explore how the number of particles in $SPA$ affects its ability to find the best combination of features and therefore its asymptotic performance on tasks. 

The approximate inference of the particle filter in $SPA$ is guided by bottom-up attention, which introduces a bias towards the most active features. This introduces a prior over the hypothesis space, favouring hypotheses with relatively few features. This is consistent with findings that people tend to make decisions based on individual features before reasoning about objects that involve more complex combinations of features \citep{farashahi2017feature, choung2017exploring}. Similarly, people find it harder to perform classification tasks as the number of relevant dimensions increases \citep{shepard1961learning}. The bottom-up attention captured in the movement step of the particle filter therefore seems to introduce a biologically plausible bias over the hypothesis space.

Top-down attention is captured during the observation step of the particle filter. This process involves evaluating different hypotheses based on their ability to predict reward \citep{mackintosh1975theory}. Interestingly \citet{radulescu2019holistic} have proposed that such a mechanism may occur in corticostriatal circuitry. \citet{radulescu2019holistic} suggest that different pools of neurons in the PFC may represent different hypotheses about the structure of the current RL task. These pools then compete via mutual lateral inhibition and this competition is biased via connections to the striatum that favour pools which lead to reward. This process parallels the observation step of $SPA$, whereby hypotheses that are more predictive of reward are more likely to be re-sampled and out-compete other hypotheses. The specific mechanism aside, the phenomenon of representing and testing multiple hypotheses during RL appears to be prevalent in human populations \citep{wilson2012inferring}.

The effectiveness of $SPA$ will naturally depend on the nature of the features or representations that it is attending too. In the human brain it has been proposed that the usefulness of selective visual feature attention decreases as you move back through the visual stream \citep{lindsay2018biological}. This is because features present later in the visual stream consist of higher-order representations that are increasingly abstract. For example one of the object categories in the multiple choice task was faces, which are known to be represented later on in the visual stream of the brain \citep{grill2018functional}. Having such a representation makes the multiple choice task easy for the brain because it only has to attend to one feature rather than a collection of low level features. This also reduces the need to consider lots of complicated hypotheses. Future work should test whether $SPA$ displays similar behaviour by testing whether its performance decreases as attention is applied to earlier convolutional layers. 

For both the multiple choice task and the object collection game, VGG-16 was used to extract features from images that were not used to train the network. In theory, using the original Imagenet dataset that it was trained on may have lead to better results in the multiple choice task. This is because VGG-16 will have already been trained to produce feature values that distinguish between the object categories in Imagenet, making it easier for $SPA$ to attend to discriminating sets of features. This hypothesis is supported by our analysis of the features produced by VGG-16 during the multiple choice task. For the combination of image categories that $SPA$ performed best on, the vectors of mean feature values were further apart in Euclidean space compared to the combination of image categories that $SPA$ performed worst on. This suggests that the more dissimilar the feature values are between the different image categories, the easier it is for $SPA$ to discriminate between them and quickly change its focus of attention.

This dependence of $SPA$ on the properties of the underlying representations that it attends to opens up several interesting avenues of future research. In particular, it would be interesting to explore whether the use of disentangled representations \citep{higgins2016beta} could further improve performance. Independent factors of variation may be easier to attend to because the informative features are separate from each other and so only simple hypotheses are required to solve the task at hand rather than complex combinations of features. Another major benefit of disentangled representations would be that the resulting attention vector would be more interpretable as each attended feature has a natural interpretation; e.g., colour or shape.

\section{Concluding Remarks}

In the present study we have presented a novel method for performing selective visual feature attention in a Deep-RL agent. Our approach, termed Selective Particle Attention ($SPA$), uses a particle filter to identify useful pre-existing features based on the task at hand. These features are then passed on to a Deep-RL algorithm to perform action selection. The particle filter incorporates bottom-up and top-down influences into the selective attention process via the movement and observation steps respectively. The movement step serves to introduce a prior over the features that are being considered so that attention is biased towards the most active features given the current input. In comparison, the movement step biases attention towards combinations of features that are most predictive of reward. Crucially these two interacting processes help to reduce the dimensionality of the learning problem in a targeted manner. This not only speeds up the efficiency of learning but also improves the agent's ability to deal with changes in the environment. Future work should explore how the nature of pre-existing representations affects the attention mechanism of $SPA$. The depth of the representation, the data used to produce them and their degree of disentanglement are all variables that may impact upon the performance of $SPA$.

\subsubsection*{Acknowledgments}

This work was funded by the UK Biotechnology and Biological Sciences Research Council (BBSRC). We thank NVIDIA for a hardware grant that provided the Graphics Processing Unit (GPU) used to run the simulations.

\urlstyle{rm}
\bibliographystyle{apalike}
\bibliography{References}

\begin{thebibliography}{}

\bibitem[Anderson et~al., 1987]{anderson1987attentional}
Anderson, D.~R., Choi, H.~P., and Lorch, E.~P. (1987).
\newblock Attentional inertia reduces distractibility during young children's
  tv viewing.
\newblock {\em Child Development}, pages 798--806.

\bibitem[Ba et~al., 2014]{ba2014multiple}
Ba, J., Mnih, V., and Kavukcuoglu, K. (2014).
\newblock Multiple object recognition with visual attention.
\newblock {\em arXiv preprint arXiv:1412.7755}.

\bibitem[Bichot et~al., 2015]{bichot2015source}
Bichot, N.~P., Heard, M.~T., DeGennaro, E.~M., and Desimone, R. (2015).
\newblock A source for feature-based attention in the prefrontal cortex.
\newblock {\em Neuron}, 88(4):832--844.

\bibitem[Bramlage and Cortese, 2020]{bramlage2020attention}
Bramlage, L. and Cortese, A. (2020).
\newblock Attention or memory? neurointerpretable agents in space and time.
\newblock {\em arXiv preprint arXiv:2007.04862}.

\bibitem[Burns and Anderson, 1993]{burns1993attentional}
Burns, J.~J. and Anderson, D.~R. (1993).
\newblock Attentional inertia and recognition memory in adult television
  viewing.
\newblock {\em Communication Research}, 20(6):777--799.

\bibitem[Choung et~al., 2017]{choung2017exploring}
Choung, O.-h., Lee, S.~W., and Jeong, Y. (2017).
\newblock Exploring feature dimensions to learn a new policy in an uninformed
  reinforcement learning task.
\newblock {\em Scientific reports}, 7(1):1--12.

\bibitem[Deng et~al., 2009]{deng2009imagenet}
Deng, J., Dong, W., Socher, R., Li, L.-J., Li, K., and Fei-Fei, L. (2009).
\newblock Imagenet: A large-scale hierarchical image database.
\newblock In {\em 2009 IEEE conference on computer vision and pattern
  recognition}, pages 248--255. Ieee.

\bibitem[Farashahi et~al., 2017]{farashahi2017feature}
Farashahi, S., Rowe, K., Aslami, Z., Lee, D., and Soltani, A. (2017).
\newblock Feature-based learning improves adaptability without compromising
  precision.
\newblock {\em Nature communications}, 8(1):1--16.

\bibitem[Fei-Fei et~al., 2006]{fei2006one}
Fei-Fei, L., Fergus, R., and Perona, P. (2006).
\newblock One-shot learning of object categories.
\newblock {\em IEEE transactions on pattern analysis and machine intelligence},
  28(4):594--611.

\bibitem[Fu et~al., 2017]{fu2017look}
Fu, J., Zheng, H., and Mei, T. (2017).
\newblock Look closer to see better: Recurrent attention convolutional neural
  network for fine-grained image recognition.
\newblock In {\em Proceedings of the IEEE conference on computer vision and
  pattern recognition}, pages 4438--4446.

\bibitem[Grill-Spector et~al., 2018]{grill2018functional}
Grill-Spector, K., Weiner, K.~S., Gomez, J., Stigliani, A., and Natu, V.~S.
  (2018).
\newblock The functional neuroanatomy of face perception: from brain
  measurements to deep neural networks.
\newblock {\em Interface Focus}, 8(4):20180013.

\bibitem[G{\"u}{\c{c}}l{\"u} and van Gerven, 2015]{gucclu2015deep}
G{\"u}{\c{c}}l{\"u}, U. and van Gerven, M.~A. (2015).
\newblock Deep neural networks reveal a gradient in the complexity of neural
  representations across the ventral stream.
\newblock {\em Journal of Neuroscience}, 35(27):10005--10014.

\bibitem[Higgins et~al., 2016]{higgins2016beta}
Higgins, I., Matthey, L., Pal, A., Burgess, C., Glorot, X., Botvinick, M.,
  Mohamed, S., and Lerchner, A. (2016).
\newblock beta-vae: Learning basic visual concepts with a constrained
  variational framework.

\bibitem[Houk and Adams, 1995]{houk199513}
Houk, J.~C. and Adams, J.~L. (1995).
\newblock 13 a model of how the basal ganglia generate and use neural signals
  that.
\newblock {\em Models of information processing in the basal ganglia}, page
  249.

\bibitem[Joel et~al., 2002]{joel2002actor}
Joel, D., Niv, Y., and Ruppin, E. (2002).
\newblock Actor--critic models of the basal ganglia: New anatomical and
  computational perspectives.
\newblock {\em Neural networks}, 15(4-6):535--547.

\bibitem[Jones and Canas, 2010]{jones2010integrating}
Jones, M. and Canas, F. (2010).
\newblock Integrating reinforcement learning with models of representation
  learning.
\newblock In {\em Proceedings of the Annual Meeting of the Cognitive Science
  Society}, volume~32.

\bibitem[Lake et~al., 2017]{lake2017building}
Lake, B.~M., Ullman, T.~D., Tenenbaum, J.~B., and Gershman, S.~J. (2017).
\newblock Building machines that learn and think like people.
\newblock {\em Behavioral and brain sciences}, 40.

\bibitem[Leong et~al., 2017]{leong2017dynamic}
Leong, Y.~C., Radulescu, A., Daniel, R., DeWoskin, V., and Niv, Y. (2017).
\newblock Dynamic interaction between reinforcement learning and attention in
  multidimensional environments.
\newblock {\em Neuron}, 93(2):451--463.

\bibitem[Lindsay, 2020]{lindsay2020attention}
Lindsay, G.~W. (2020).
\newblock Attention in psychology, neuroscience, and machine learning.
\newblock {\em Frontiers in Computational Neuroscience}, 14:29.

\bibitem[Lindsay and Miller, 2018]{lindsay2018biological}
Lindsay, G.~W. and Miller, K.~D. (2018).
\newblock How biological attention mechanisms improve task performance in a
  large-scale visual system model.
\newblock {\em ELife}, 7:e38105.

\bibitem[Longman et~al., 2014]{longman2014attentional}
Longman, C.~S., Lavric, A., Munteanu, C., and Monsell, S. (2014).
\newblock Attentional inertia and delayed orienting of spatial attention in
  task-switching.
\newblock {\em Journal of Experimental Psychology: Human Perception and
  Performance}, 40(4):1580.

\bibitem[Luo et~al., 2020]{luo2020costs}
Luo, X., Roads, B.~D., and Love, B.~C. (2020).
\newblock The costs and benefits of goal-directed attention in deep
  convolutional neural networks.
\newblock {\em arXiv preprint arXiv:2002.02342}.

\bibitem[Mackintosh, 1975]{mackintosh1975theory}
Mackintosh, N.~J. (1975).
\newblock A theory of attention: Variations in the associability of stimuli
  with reinforcement.
\newblock {\em Psychological review}, 82(4):276.

\bibitem[Maia, 2009]{maia2009reinforcement}
Maia, T.~V. (2009).
\newblock Reinforcement learning, conditioning, and the brain: Successes and
  challenges.
\newblock {\em Cognitive, Affective, \& Behavioral Neuroscience},
  9(4):343--364.

\bibitem[Manchin et~al., 2019]{manchin2019reinforcement}
Manchin, A., Abbasnejad, E., and van~den Hengel, A. (2019).
\newblock Reinforcement learning with attention that works: A self-supervised
  approach.
\newblock In {\em International Conference on Neural Information Processing},
  pages 223--230. Springer.

\bibitem[Miller and Cohen, 2001]{miller2001integrative}
Miller, E.~K. and Cohen, J.~D. (2001).
\newblock An integrative theory of prefrontal cortex function.
\newblock {\em Annual review of neuroscience}, 24(1):167--202.

\bibitem[Mnih et~al., 2016]{mnih2016asynchronous}
Mnih, V., Badia, A.~P., Mirza, M., Graves, A., Lillicrap, T., Harley, T.,
  Silver, D., and Kavukcuoglu, K. (2016).
\newblock Asynchronous methods for deep reinforcement learning.
\newblock In {\em International conference on machine learning}, pages
  1928--1937.

\bibitem[Mnih et~al., 2014]{mnih2014recurrent}
Mnih, V., Heess, N., Graves, A., et~al. (2014).
\newblock Recurrent models of visual attention.
\newblock In {\em Advances in neural information processing systems}, pages
  2204--2212.

\bibitem[Mnih et~al., 2015]{mnih2015human}
Mnih, V., Kavukcuoglu, K., Silver, D., Rusu, A.~A., Veness, J., Bellemare,
  M.~G., Graves, A., Riedmiller, M., Fidjeland, A.~K., Ostrovski, G., et~al.
  (2015).
\newblock Human-level control through deep reinforcement learning.
\newblock {\em nature}, 518(7540):529--533.

\bibitem[Mott et~al., 2019]{mott2019towards}
Mott, A., Zoran, D., Chrzanowski, M., Wierstra, D., and Rezende, D.~J. (2019).
\newblock Towards interpretable reinforcement learning using attention
  augmented agents.
\newblock In {\em Advances in Neural Information Processing Systems}, pages
  12329--12338.

\bibitem[Niv et~al., 2015]{niv2015reinforcement}
Niv, Y., Daniel, R., Geana, A., Gershman, S.~J., Leong, Y.~C., Radulescu, A.,
  and Wilson, R.~C. (2015).
\newblock Reinforcement learning in multidimensional environments relies on
  attention mechanisms.
\newblock {\em Journal of Neuroscience}, 35(21):8145--8157.

\bibitem[Paneri and Gregoriou, 2017]{paneri2017top}
Paneri, S. and Gregoriou, G.~G. (2017).
\newblock Top-down control of visual attention by the prefrontal cortex.
  functional specialization and long-range interactions.
\newblock {\em Frontiers in neuroscience}, 11:545.

\bibitem[Radulescu et~al., 2019a]{radulescu2019holistic}
Radulescu, A., Niv, Y., and Ballard, I. (2019a).
\newblock Holistic reinforcement learning: the role of structure and attention.
\newblock {\em Trends in cognitive sciences}.

\bibitem[Radulescu et~al., 2019b]{radulescuparticle}
Radulescu, A., Niv, Y., and Daw, N.~D. (2019b).
\newblock A particle filtering account of selective attention during learning.
\newblock {\em 2019 Conference on Cognitive Computational Neuroscience}.

\bibitem[Richards and Anderson, 2004]{richards2004attentional}
Richards, J.~E. and Anderson, D.~R. (2004).
\newblock Attentional inertia in children's extended looking at television.
\newblock In {\em Advances in child development and behavior}, volume~32, pages
  163--212. Elsevier.

\bibitem[Rossi and Paradiso, 1995]{rossi1995feature}
Rossi, A.~F. and Paradiso, M.~A. (1995).
\newblock Feature-specific effects of selective visual attention.
\newblock {\em Vision research}, 35(5):621--634.

\bibitem[Saenz et~al., 2002]{saenz2002global}
Saenz, M., Buracas, G.~T., and Boynton, G.~M. (2002).
\newblock Global effects of feature-based attention in human visual cortex.
\newblock {\em Nature neuroscience}, 5(7):631--632.

\bibitem[Schrimpf et~al., 2018]{schrimpf2018brain}
Schrimpf, M., Kubilius, J., Hong, H., Majaj, N.~J., Rajalingham, R., Issa,
  E.~B., Kar, K., Bashivan, P., Prescott-Roy, J., Schmidt, K., et~al. (2018).
\newblock Brain-score: Which artificial neural network for object recognition
  is most brain-like?
\newblock {\em BioRxiv}, page 407007.

\bibitem[Schultz, 1998]{schultz1998predictive}
Schultz, W. (1998).
\newblock Predictive reward signal of dopamine neurons.
\newblock {\em Journal of neurophysiology}, 80(1):1--27.

\bibitem[Schultz et~al., 1997]{schultz1997neural}
Schultz, W., Dayan, P., and Montague, P.~R. (1997).
\newblock A neural substrate of prediction and reward.
\newblock {\em Science}, 275(5306):1593--1599.

\bibitem[Schyns et~al., 1998]{schyns1998development}
Schyns, P.~G., Goldstone, R.~L., and Thibaut, J.-P. (1998).
\newblock The development of features in object concepts.
\newblock {\em Behavioral and brain Sciences}, 21(1):1--17.

\bibitem[Setlow et~al., 2003]{setlow2003neural}
Setlow, B., Schoenbaum, G., and Gallagher, M. (2003).
\newblock Neural encoding in ventral striatum during olfactory discrimination
  learning.
\newblock {\em Neuron}, 38(4):625--636.

\bibitem[Shepard et~al., 1961]{shepard1961learning}
Shepard, R.~N., Hovland, C.~I., and Jenkins, H.~M. (1961).
\newblock Learning and memorization of classifications.
\newblock {\em Psychological monographs: General and applied}, 75(13):1.

\bibitem[Simonyan and Zisserman, 2014]{simonyan2014very}
Simonyan, K. and Zisserman, A. (2014).
\newblock Very deep convolutional networks for large-scale image recognition.
\newblock {\em arXiv preprint arXiv:1409.1556}.

\bibitem[Sorokin et~al., 2015]{sorokin2015deep}
Sorokin, I., Seleznev, A., Pavlov, M., Fedorov, A., and Ignateva, A. (2015).
\newblock Deep attention recurrent q-network.
\newblock {\em arXiv preprint arXiv:1512.01693}.

\bibitem[Treue, 2003]{treue2003visual}
Treue, S. (2003).
\newblock Visual attention: the where, what, how and why of saliency.
\newblock {\em Current opinion in neurobiology}, 13(4):428--432.

\bibitem[Treue and Trujillo, 1999]{treue1999feature}
Treue, S. and Trujillo, J. C.~M. (1999).
\newblock Feature-based attention influences motion processing gain in macaque
  visual cortex.
\newblock {\em Nature}, 399(6736):575--579.

\bibitem[Wilson and Niv, 2012]{wilson2012inferring}
Wilson, R.~C. and Niv, Y. (2012).
\newblock Inferring relevance in a changing world.
\newblock {\em Frontiers in human neuroscience}, 5:189.

\bibitem[Wolfe and Horowitz, 2004]{wolfe2004attributes}
Wolfe, J.~M. and Horowitz, T.~S. (2004).
\newblock What attributes guide the deployment of visual attention and how do
  they do it?
\newblock {\em Nature reviews neuroscience}, 5(6):495--501.

\bibitem[Xiao et~al., 2015]{xiao2015application}
Xiao, T., Xu, Y., Yang, K., Zhang, J., Peng, Y., and Zhang, Z. (2015).
\newblock The application of two-level attention models in deep convolutional
  neural network for fine-grained image classification.
\newblock In {\em Proceedings of the IEEE conference on computer vision and
  pattern recognition}, pages 842--850.

\bibitem[Xu et~al., 2015]{xu2015show}
Xu, K., Ba, J., Kiros, R., Cho, K., Courville, A., Salakhudinov, R., Zemel, R.,
  and Bengio, Y. (2015).
\newblock Show, attend and tell: Neural image caption generation with visual
  attention.
\newblock In {\em International conference on machine learning}, pages
  2048--2057.

\bibitem[Yamins and DiCarlo, 2016]{yamins2016using}
Yamins, D.~L. and DiCarlo, J.~J. (2016).
\newblock Using goal-driven deep learning models to understand sensory cortex.
\newblock {\em Nature neuroscience}, 19(3):356--365.

\bibitem[Zambaldi et~al., 2018]{zambaldi2018relational}
Zambaldi, V., Raposo, D., Santoro, A., Bapst, V., Li, Y., Babuschkin, I.,
  Tuyls, K., Reichert, D., Lillicrap, T., Lockhart, E., et~al. (2018).
\newblock Relational deep reinforcement learning.
\newblock {\em arXiv preprint arXiv:1806.01830}.

\end{thebibliography}
\end{document}